\documentclass[twocolumn,superscriptaddress,nobibnotes,nofootinbib]{revtex4-2}
\usepackage[utf8]{inputenc}
\usepackage{amsmath}
\usepackage[english]{babel}
\usepackage[dvipsnames]{xcolor}
\usepackage{amssymb}
\usepackage{textcomp}
\usepackage{graphicx}
\usepackage{placeins}
\usepackage{textgreek}
\usepackage{upgreek}
\usepackage{latexsym}
\usepackage{braket}
\usepackage[nolist]{acronym}
\usepackage{siunitx}
\usepackage{xpatch}
\usepackage{dsfont}
\usepackage{xcolor}
\usepackage{hyperref}
\hypersetup{linktoc=all, hyperindex=true,colorlinks, linkcolor={red!50!black},citecolor={blue!50!black},urlcolor={blue!80!black}}

\makeatletter

\def\l@subsection#1#2{}
\def\l@subsubsection#1#2{}
\patchcmd{\@ssect@ltx}
    {\addcontentsline{toc}{#1}{\protect\numberline{}#8}}
    {}
    {}
    {}
\makeatother

\DeclareMathOperator{\diag}{diag}
\DeclareMathOperator{\tr}{tr}
\begin{document}
%
\newacro{lqg}[LQG]{linear quadratic gaussian regulator}
\newacro{lqr}[LQR]{linear quadratic regulator}

\newcommand{\bop}{\ensuremath{b}}
\newcommand{\bdag}{\ensuremath{b^{\dagger}}}
\newcommand{\aop}{\ensuremath{a}}
\newcommand{\adag}{\ensuremath{a^{\dagger}}}
\newcommand{\dd}[1]{\ensuremath{\mathrm{d}#1\,}}
\newcommand{\ii}{\ensuremath{\mathrm{i}}}
\newcommand{\ee}{\ensuremath{\mathrm{e}}}
\newcommand{\ddd}[1]{\ensuremath{\mathrm{d}^3#1\,}}
\newcommand{\mean}[1]{\ensuremath{\langle  #1 \rangle}}
%
\title{Real-time optimal quantum control of mechanical motion at room temperature}

\newcommand{\vcq}{University of Vienna, Faculty of Physics, Vienna Center for Quantum Science and Technology (VCQ), 1090 Vienna, Austria}

\newcommand{\iqoqi}{Institute for Quantum Optics and Quantum Information (IQOQI), Austrian Academy of Sciences, 1090 Vienna, Austria.}

\newcommand{\acin}{Automation and Control Institute (ACIN), TU Wien, 1040 Vienna, Austria}

\newcommand{\ait}{Austrian Institute of Technology (AIT), Center for Vision, Automation \& Control 1040, Vienna, Austria}

\newcommand{\stu}{Institute for Functional Matter and Quantum Technologies (FMQ) and Center for Integrated Quantum Science and Technology (IQST), University of Stuttgart, 70569 Stuttgart, Germany}

%
\author{Lorenzo Magrini}
\email{lorenzo.magrini@univie.ac.at}
\affiliation{\vcq}
\author{Philipp Rosenzweig}
\affiliation{\acin}
\author{Constanze Bach}
\affiliation{\vcq}
\author{Andreas Deutschmann-Olek}
\affiliation{\acin}
\author{Sebastian G.\ Hofer}
\affiliation{\vcq}
\author{Sungkun Hong}
\affiliation{\stu}
\author{Nikolai Kiesel}
\affiliation{\vcq}
\author{Andreas Kugi}
\affiliation{\acin}
\affiliation{\ait}
\author{Markus Aspelmeyer}
\email{markus.aspelmeyer@univie.ac.at}
\affiliation{\vcq}
\affiliation{\iqoqi}

\begin{abstract}
The ability to accurately control the dynamics of physical systems by measurement and feedback is a pillar of modern engineering~\cite{aastrom2013computer}. Today, the increasing demand for applied quantum technologies requires to adapt this level of control to individual quantum systems~\cite{Geremia2003, Glaser2015}. Achieving this in an optimal way is a challenging task that relies on both quantum-limited measurements and specifically tailored algorithms for state estimation and feedback~\cite{Wieseman2010}. Successful implementations thus far include experiments on the level of optical and atomic systems~\cite{Sayrin2011, Yonezawa2012, Martinez2018}. Here we demonstrate real-time optimal control of the quantum trajectory~\cite{Carmichael1993} of an optically trapped nanoparticle. We combine confocal position sensing close to the Heisenberg limit with optimal state estimation via Kalman filtering to track the particle motion in phase space in real time with a position uncertainty of 1.3 times the zero point fluctuation. Optimal feedback allows us to stabilize the quantum harmonic oscillator to a mean occupation of $n=0.56\pm0.02$ quanta, realizing quantum ground state cooling from room temperature. Our work establishes quantum Kalman filtering as a method to achieve quantum control of mechanical motion, with potential implications for sensing on all scales. In combination with levitation, this paves the way to full-scale control over the wavepacket dynamics of solid-state macroscopic quantum objects in linear and nonlinear systems.
\end{abstract}

\flushbottom
\maketitle

\thispagestyle{empty}


The Kalman filter is an iterative real-time state estimation algorithm that combines measurement records with a mathematical description of the system dynamics. At each time step, it provides a state estimate that is conditioned on the knowledge acquired from earlier observations~\cite{kalman1960}. This \textit{conditional state} can then serve as the basis for feedback control methods that steer the system and stabilize it in a desired target state~\cite{kalman1960lqr}. For Gaussian systems, the Kalman filter is optimal in a mean-square-error sense. As many physical systems can be approximated by Gaussian dynamics it is being used in a broad variety of applications ranging from bio-medical signal processing~\cite{Sittig1992} over navigation~\cite{bar2004estimation} to mechanical sensing~\cite{Ruppert2016}. In particular for the last case, high-precision experiments employing mechanical sensors are now approaching a regime in which quantum effects of the object itself become relevant~\cite{Rossi2018,Rossi2019}. Any estimation or control approach therefore has to incorporate a full quantum description~\cite{Wieseman2010}. 
In analogy with the classical case, the dynamics of an open quantum system undergoing continuous measurement can be generally understood as a non-linear quantum filtering problem, giving rise to the concept of \textit{conditional quantum states}. It was shown by Belavkin~\cite{Belavkin1980} that for Gaussian systems the quantum filter reduces to the classical Kalman-filter form. Critically, however, quantum mechanics places restrictions on the underlying physical model, in particular to reflect the intrusive nature of the measurement. The challenge in realizing real-time (optimal) quantum control is then two-fold: First, the measurement process has to be quantum limited, i.e., imprecision and backaction of the measurement must saturate the Heisenberg uncertainty relation. This is achieved only for a high detection efficiency and if the decoherence of the system is dominated by the quantum backaction of the measurement process.
Second, quantum filtering has to be implemented in real time and connected to a feedback architecture that allows to stabilize the desired quantum state. 
For mechanical devices, these requirements have thus far only been realized independently in separate experiments. 
In a cryogenic environment, ground-state feedback cooling~\cite{Rossi2018} and offline quantum filtering~\cite{Rossi2019} were demonstrated for a micromechanical resonator. In a regime driven by thermal forces, Kalman filtering was implemented for classical feedback on a gram-scale mirror~\cite{Iwasawa2013}, offline state estimation of micromechanical motion~\cite{Wieczorek2015}, and real-time state estimation and feedback of nanomechanical systems~\cite{Setter2018, Liao2019}. In a backaction dominated regime, feedback was used to cool mechanical motion close to the quantum ground state with suspended nanobeams~\cite{Suhdir2017} and levitated nanoparticles~\cite{Tebbenjohanns2020,Kamba2020}.
As of yet, optimal control at the quantum level has not been achieved. Our work combines all relevant elements in a single experiment, specifically optimal state estimation based on near-Heisenberg limited measurement sensitivity at room temperature with optimal control of the quantum trajectory. Consequently, we can stabilize the unconditional quantum state of a levitated nanoparticle to a position uncertainty of 1.3 times the ground state extension. This contrasts cavity-based cooling schemes for levitated nanoparticles~\cite{Windey2019,Delic2019,DelosRiosSommer2021} that also achieve ground-state cooling\cite{Delic2020} but without requiring quantum-limited readout sensitivity. In comparison, real-time optimal control as presented here avoids the overhead of cavity stabilization and can tolerate colored environmental noise by including it directly in the state-space model~\cite{Wieczorek2015}.

\section*{Quantum-limited measurement}

\begin{figure}
    \centering
    \includegraphics[scale=1]{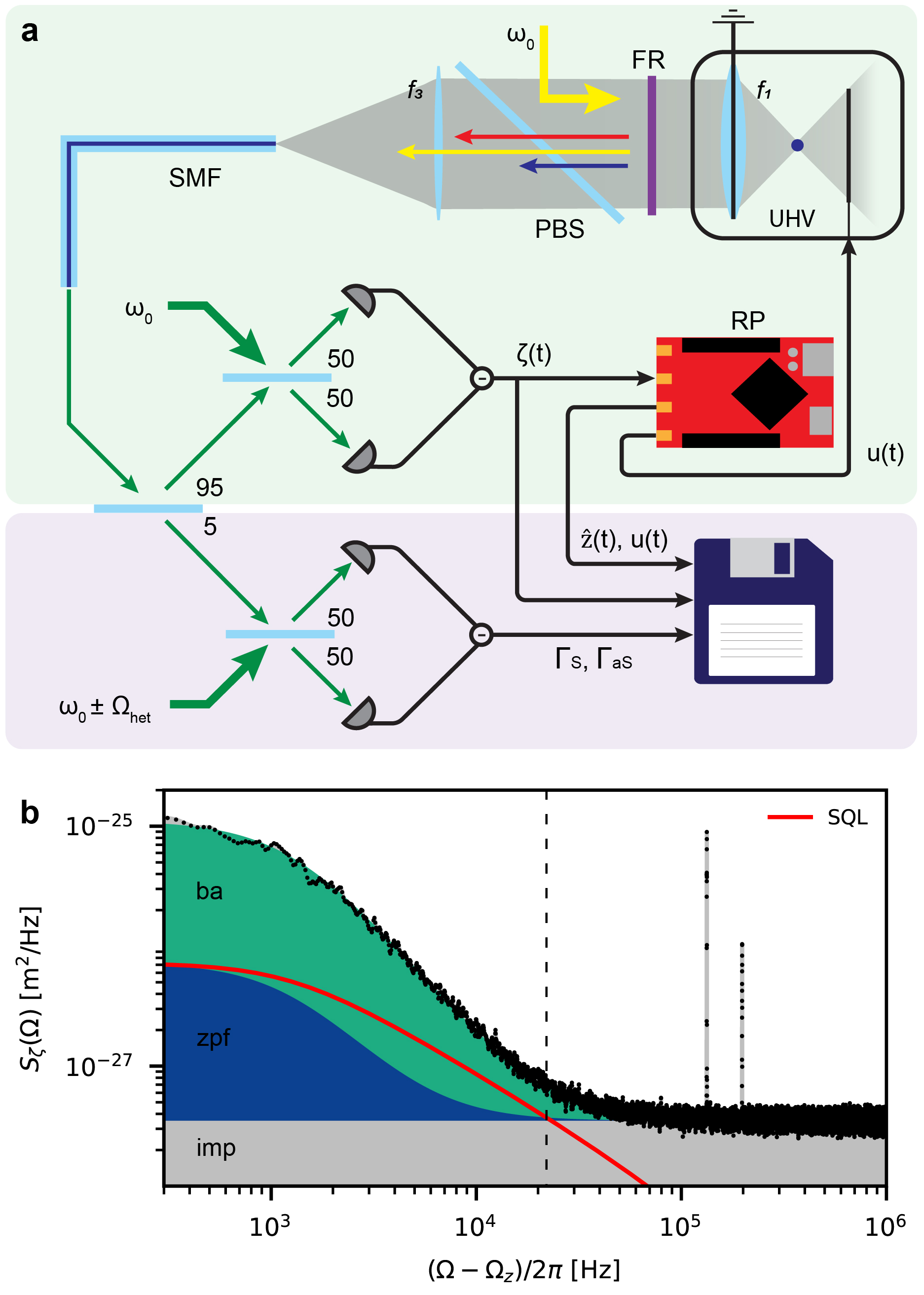}
    \caption{\textbf{Experimental setup}. \textbf{a}, Scheme of the experimental setup. The particle is trapped in an optical tweezer (laser frequency: $\omega_0$), and oscillates in an utra-high vacuum (UHV), along the $z$ direction, at a frequency of $\Omega_z/2\pi = 104\, \mathrm{kHz}$. The backscattered light is collected by the tweezer objective lens ($f_1$), separated from the tweezer light by the combination of a faraday rotator (FR) and polarizing beam-splitter (PBS) and spatially filtered by focusing ($f_3$) onto a single mode fiber (SMF) in a confocal arrangement. It is then split into two paths: an in-loop homodyne detection and an out-of-loop heterodyne detection. The homodyne detection is used for the efficient position measurement ($\zeta(t)$), and is directed to the Red-Pitaya (RP) board, where the LQG is implemented in real time. Both the state estimate ($\mathbf{\hat{z}}(t)$) the and control signal  ($u(t)$) can be recorded. The control signal is applied to the electrode in the vacuum chamber. The heterodyne detection (local oscillator at a frequency of $\omega_0 \pm \Omega_{\mathrm{het}}$) employs only 5\% of the light and performs an out-of-loop measurement of the particle's energy via Raman scattering thermometry by measurement of the ratio of the Stokes and anti-Stokes scattering rates ($\Gamma_\mathrm{S}$, $\Gamma_\mathrm{aS}$). \textbf{b}, Contributions to the measured position power spectral density by the measurement imprecision (imp), the measurement backaction (ba), and the mechanical quantum fluctuations (zpf) in the homodyne detection, at a control gain of $g_\mathrm{fb}/2\pi = 2 \, \mathrm{kHz}$ and occupation $\left\langle n\right\rangle = 8.3\pm 0.09$. The dashed line indicates the frequency ($\sim \Omega_z + 2\pi\cdot22\ \mathrm{kHz}$) at which imprecision and backaction contribute equally to the total added noise. Here the measured noise is only a fator 1.76 above the SQL (red line).}
    \label{fig:1}
\end{figure}

We use an optical tweezer (NA = 0.95, $\lambda_0$ = \SI{1064}{\nano\metre}, power $\approx $\SI{300}{\milli\watt}, linearly polarized) to trap a silica nanosphere of $\SI{71.5}{\nano\metre}$ radius ($\approx 2.8\times 10^{-18} \mathrm{kg}$) in ultra-high vacuum (Figure \ref{fig:1}a). The particle oscillates at frequencies of $\Omega_z/2\pi=\SI{104}{\kilo\hertz}$, $\Omega_y/2\pi=\SI{236}{\kilo\hertz}$ and $\Omega_x/2\pi=\SI{305}{\kilo\hertz}$, where we use the trapping beam to define a coordinate system with $z$ along the beam axis and $x$ and $y$ parallel and perpendicular to its polarization, respectively. The motion in the $x$- and $y$-direction is stabilized by an independent parametric feedback to occupations of about $10^3$, allowing us to suppress any effect due to thermal nonlinearties or measurement cross-coupling~\cite{Gieseler2013, methods}.
Most trapped particles carry excess charges, which allows us to apply a calibrated force through an external electric field. In our case, we control the $z$-motion by a voltage applied to an electrode in front of the grounded tweezer objective~\cite{Frimmer2017}.
The position of the particle is encoded in the optical phase of the scattered tweezer light, which is collected and measured by optical homodyning. Note that the position information contained in the scattered light is not uniformly distributed~\cite{Tebbenjohanns2019_detection, Seberson2019}. For the $z$-direction, almost all information is carried by the backscattered photons, which is why we restrict ourselves to backplane detection using a fiber-based confocal microscope~\cite{Vamivakas2007}. Here the collected light is spatially filtered by a single-mode fiber, which suppresses contributions from stray light by almost a factor $10^3$ while maximizing the overlap between the spatial modes of the scattering dipole and the fiber ($\eta_\mathrm{m}=0.71$)~\cite{methods}.
Our measurement operates close to the quantum limit. In the ideal case, imprecision and backaction noise of the measurement
saturate the Heisenberg uncertainty relation $\sqrt{S_z^\mathrm{I}(\Omega)S_F^{\mathrm{ba}}(\Omega)}=\hbar$ for all frequencies $\Omega$ ($S_{z,F}(\Omega)$: one-sided noise power spectral densities of position (z) and force (F)~\cite{methods}). Losses degrade this performance: experimental losses in the detection channel ($\eta_\mathrm{d}$) increase the imprecision noise to $S_z^{\mathrm {imp}}=S_z^\mathrm{I}/\eta_\mathrm{d}$, while additional environmental interactions, for example scattering of gas molecules, increase the total force noise to $S_F^{\mathrm{tot}}=S_F^{\mathrm{\mathrm{ba}}}/\eta_\mathrm{e}$. This results in $\sqrt{S_z^{\mathrm{imp}}(\Omega)S_F^{\mathrm{tot}}(\Omega)}=\hbar/\sqrt{\eta}$, where $\eta=\eta_\mathrm{d}\eta_\mathrm{e}$ amounts to an effective collection efficiency of the overall phase-space information available from the system. In our case, the efficient and low-noise confocal detection scheme results in a displacement sensitivity of $\sqrt{S_z^{\mathrm{imp}}} = 2.0\times10^{-14}~\mathrm{m/\sqrt{\mathrm{Hz}}}$, allowing us to resolve displacements of the size of the zero-point motion of the particle ($z_{\mathrm{zpf}}=\sqrt{\hbar/(2m\Omega_z)}$) at a rate of $\Gamma_{\mathrm{meas}} = z_{{\mathrm{zpf}}}^2/2S_z^{{\mathrm{imp}}} = 2\pi\cdot 6.6\ \mathrm{kHz}$~\cite{Clerk2010}. By performing re-heating measurements at different background pressures, we can directly determine the decoherence rates of the particle due to backaction, $\Gamma_{{\mathrm{ba}}} = 2\pi\cdot 18.8\ \mathrm{kHz}$, and due to residual gas molecules, $\Gamma_{{\mathrm{th}}} = 2\pi\cdot 0.6\ \mathrm{kHz}$ at the minimal operating pressure of $9.2\times 10^{-9}\mathrm{mbar}$, providing us with a quantum cooperativity of $C_q = \Gamma_{{\mathrm{ba}}}/ \Gamma_{{\mathrm{th}}} = 30$~\cite{methods}. The resulting information collection efficiency~\cite{Clerk2010} $\eta = \Gamma_{{\mathrm{meas}}}/\left(\Gamma_{\mathrm{ba}} + \Gamma_{\mathrm{th}} \right) = 0.34$ is consistent with the value obtained from the independently measured loss contributions in the experimental setup~\cite{methods}. This yields an imprecision--backaction product of
$\hbar/\sqrt{\eta} = 1.7\hbar$, which is less than a factor of 2 from its fundamental limit, and more than one order of magnitude better than previously shown for mechanical systems at room temperature~\cite{Abbott2009, Bushev2013, Tebbenjohanns2020, Kamba2020}. Note that this also enables measurements close to the standard quantum limit (SQL), where the effects of imprecision and backaction force noise on the displacement spectrum are equal. Figure \ref{fig:1}b shows the different noise contributions for a measurement performed at moderate feedback gain, where a sensitivity of 1.76 times the SQL is reached at frequencies of $\sim22\,\mathrm{kHz}$ above resonance.  

\section*{Optimal quantum control}
The idea of optimal feedback is to find a control input that renders the closed-loop system stable and optimizes a pre-defined cost function. In our case, the goal is to minimize the particle's energy. This task can be broken down into two steps: an estimation step to provide an optimal estimate of the system's quantum state in real time, here in the form of a Kalman filter; and a control step that computes the optimal feedback, here in the form of a linear--quadratic regulator (LQR). Both steps require an adequate mathematical model of the experimental setup, and together form the so-called linear--quadratic--Gaussian (LQG) control problem. To this end, we define a quantum stochastic model that allows us to construct the dynamical equations for the conditional quantum state $\hat{\rho}$. We model the levitated particle as a one-dimensional quantum harmonic oscillator coupling to two environments, the electromagnetic field in the vacuum state and the residual gas in a thermal state. Both environments are treated in a Markovian approximation, which means they effectively act as Gaussian white noise sources. By measuring the electromagnetic field we realize a (continuous) measurement of the particle position.
As under this model the system state is Gaussian at all times, $\hat{\rho}$ is fully characterized by the first two moments of the state vector ${\mathbf{z}} = [z,p]^{\mathrm{T}}$ ($z$ and $p$ being the particle's position and momentum operators in the $z$-direction), given by $\hat{\mathbf{z}}(t)=\tr(\mathbf{z}\hat{\rho}(t) )$ and $\hat{\mathbf{\Sigma}}(t) = \mathrm{Re}[\tr({\mathbf{z}}{\mathbf{z}}^{\mathrm{T}}\hat{\rho}(t) )] - \hat{\mathbf{z}}(t)\hat{\mathbf{z}}(t)^{\mathrm{T}}$. Here we follow the notation where the $\hat{}$ -symbol refers to the quantities of the conditional state.
The corresponding equations of motion for $\hat{\mathbf{z}}$ and $\hat{\mathbf{\Sigma}}$ are then equivalent to the classical Kalman--Bucy filter~\cite{Belavkin1980, Doherty1999, methods}, which takes the noisy measurement signal $\zeta(t)$ as an input.
\begin{figure}
    \centering
    \includegraphics[=1]{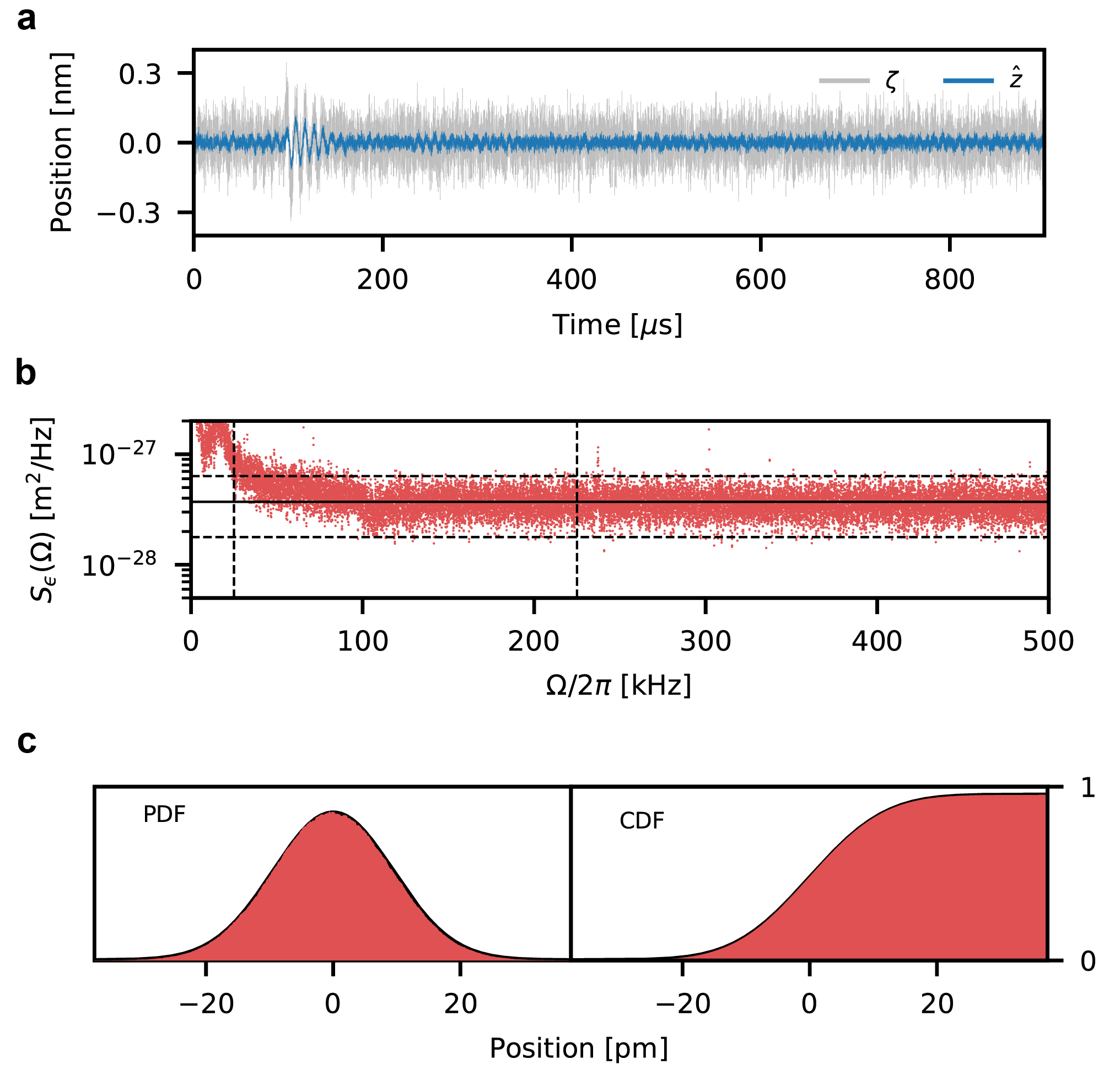}
    \caption{\textbf{Kalman filter and verification}. \textbf{a}, Time trace of the measurement (gray) and estimation (blue) sequences at $g_\mathrm{fb}/2\pi = 16\, \mathrm{kHz} $, $n = 1.68\pm 0.09$. At around $t=100\,\upmu\mathrm{s}$, a (rare, $\sim 10p_{\mathrm{zpf}}$) disturbance to the particle is highlighted by the filter. \textbf{b}, Power spectral density of the innovation sequence. Horizontal lines indicate the white noise model (solid) and the $95\%$ confidence region of the expected $\chi^2$ distribution (dashed)~\cite{Wieczorek2015}. The low frequency phase noise ($< 25\,\mathrm{kHz}$) and the narrow noise peaks due to residual $x$ -- and $y$ -- modes coupling ($> 225\,\mathrm{kHz}$) are not considered in our noise model. \textbf{c}, Experimental probability density function (PDF) and cumulative density function (CDF) of a $10\,\mathrm{ms}$ innovation sequence. A 4th order $f_c = 10\,\mathrm{kHz}$ highpass filter is used to reduce the low frequency contributions that are not considered in our model. The black lines are Gaussian fits to the data.}
    \label{fig:2}
\end{figure}    
The particle's motion is controlled by a control input $u(t)$, which defines the feedback force that is applied to the particle via an external electric field: $F_\mathrm{fb} = qE_\mathrm{fb}(t) = \hbar u(t)/z_\mathrm{zpf}$ ($q$: the charge of the particle, $E_\mathrm{fb}(t)$: the electric field.)
In order to find the optimal control input $u(t)=-{\mathbf{k}^{\mathrm{T}}(t)}{\hat{\mathbf{z}}(t)}$ ($\mathbf{k}^{\mathrm{T}}(t)$ being the feedback vector) that minimizes the particle's energy, we solve the (deterministic) LQR problem~\cite{kalman1960lqr,Doherty1999,methods}. The solution depends on the control effort, which can be parametrized by the feedback gain $g_{\mathrm{fb}}$. Adjusting this degree of freedom allows us to shape the closed-loop dynamics and steer the particle into the desired thermal state. The corresponding closed-loop covariance matrix of $\mathbf{z}$ is given by $\mathbf{\Sigma}(t)=\hat{\mathbf{\Sigma}}(t)+\langle \mathbf{\hat{z}}(t)\mathbf{\hat{z}}(t)^\mathrm{T}\rangle_{\mathrm{cl}}$, where $\langle \cdot\rangle_{\mathrm{cl}}$ denotes the expectation value with respect to the classical stochastic process induced by the measurement. In the long term limit ($t \gg 1/\Gamma_\mathrm{meas}$), both $\mathbf{\Sigma}(t)$ and $\mathbf{\hat{\Sigma}}(t)$ converge to a steady state, which we denote by $\mathbf{\Sigma}^\mathrm{ss}$ and $\mathbf{\hat{\Sigma}}^\mathrm{ss}$ respectively. Then $\mathbf{\hat{\Sigma}}^\mathrm{ss}$ can be obtained by solving the stationary Riccati equation.
Finally, we combine the stationary LQR and Kalman filter into a single time-discrete transfer function that solves the optimal quantum feedback problem in real time. It is implemented as a digital filter with a sampling time of $T_s = \SI{32}{\nano\second}$ in a \textsf{Red Pitaya} board equipped with a Xilinx Zynq 7010 FPGA.
A key element of optimal estimation and control is the accurate mathematical description of the experimental setup including external noise processes, which relies on a careful calibration of the position readout. We calibrate our readout using Raman sideband thermometry from an out-of-loop heterodyne detection, which provides an absolute energy measurement that is compared to the simultaneously recorded homodyne position measurement. To avoid any possible distortion in the closed-loop position detection that may result in noise squashing~\cite{Poggio2007}, we perform the calibration at low feedback gains~\cite{methods}.
This allows us to quantify all relevant noise processes and to calibrate the feedback force applied via the electrodes~\cite{methods}. We ensure the accuracy of the conditional state computed by the Kalman filter by performing a thorough model verification. This is a crucial aspect, in particular because the dynamical equations for $\hat{\mathbf{\Sigma}}$ do not depend on the measurements but only on the model. Verification is done by computing the innovation sequence $\epsilon(t) = \zeta(t)-\hat{z}(t)$, which describes the difference between the position predicted by the Kalman filter $\hat{z}(t)$ and the actual measurement outcome $\zeta(t)$. For an optimally working filter, $\epsilon$ is a Gaussian zero-mean white noise process. We confirm this to be the case for our experiment, (see Figure \ref{fig:2}b-c).

\section*{Results}
\begin{figure*}[!ht]
    \includegraphics[scale =1]{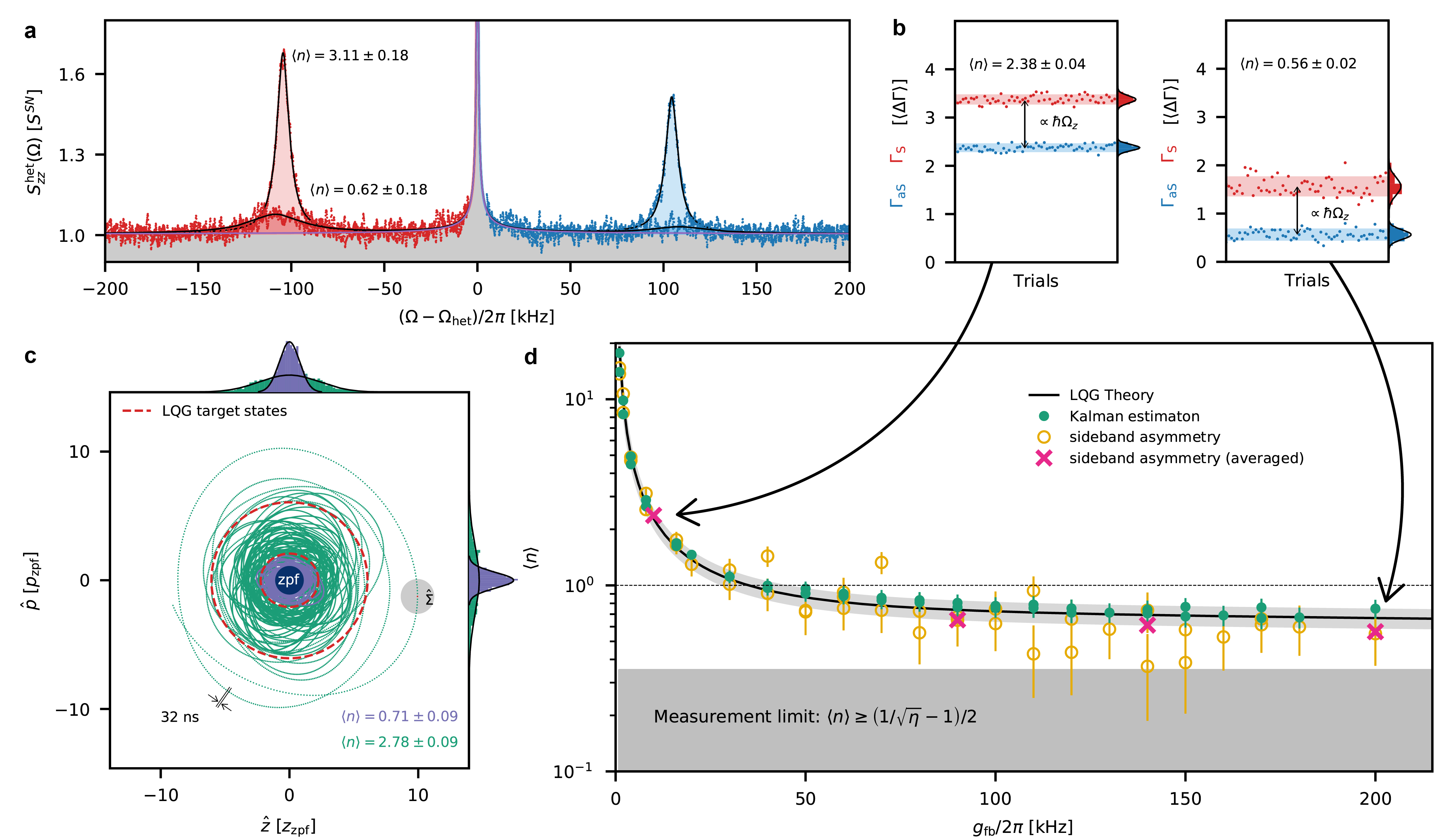}
    \caption{\textbf{Quantum optimal control}. \textbf{a}, Heterodyne power spectral density at $g_\mathrm{fb}/2\pi = 8 \,\mathrm{kHz}$ (large narrow peaks) and $g_\mathrm{fb}/2\pi = 110 \,\mathrm{kHz}$ (small broad peaks), where we distinguish the spectral contributions from Stokes (red) and anti-Stokes (blue) scattering.  The asymmetry of the peaks is a signature of the quantization of the energy levels of the harmonic oscillator. \textbf{b}, Statistical fluctuations of Stokes (red) and anti-Stokes (blue) scattering rates at $g_\mathrm{fb}/2\pi = 10 \,\mathrm{kHz}$ and $g_\mathrm{fb}/2\pi = 200 \,\mathrm{kHz}$. Each point is evaluated by integrating a single PSD as shown in \textbf{a}, and normalizing by the average value of their difference over all of the measurements ($\left\langle \Delta\Gamma \right\rangle = \left\langle \Gamma_{\mathrm{S}}-\Gamma_{\mathrm{aS}} \right\rangle)$. \textbf{c}, Phase space plot of the quantum trajectory of the particle at the steady state, for $g_\mathrm{fb}/2\pi = 8 \,\mathrm{kHz}$ (green), $g_\mathrm{fb}/2\pi = 110  \,\mathrm{kHz}$ (purple) and the corresponding solutions of the LQG closed-loop system (red dashed). Both traces display about $750\,\mathrm{\upmu s}$ of evolution. Highlighted is the uncertainty given by the steady-state conditional covariance matrix $\hat{\mathbf{\Sigma}}^\mathrm{ss}$ as given by the Kalman filter. For comparison, we show the phase space volume occupied by the zero-point fluctuations in dark blue. Here the data is filtered with a high-order bandpass (\SIrange[range-units=single]{25}{225}{\kilo\hertz}), attenuating the contributions of the noise sources at high and low frequencies that are not considered by the model. \textbf{d}, Occupation at different feedback gains as estimated by the Kalman filter (green dots) and independently measured by heterodyne asymmetry (yellow circles). The magenta crosses show the four points at which 60 repeated measurements were performed for reduction of statistical fluctuations as in \textbf{b}. Error bars represent the standard deviation of the measured value. The solid line is the analytic closed-loop solution of the LQG, showing the expected occupancy given by our experimental parameters and their uncertainties. The gray area shows the cooling limit set by the efficiency of our measurement.}
    \label{fig:3}
\end{figure*}
The closed-loop dynamics can be influenced by adjusting the feedback gain $g_\mathrm{fb}$. At each gain setting, we record the measurement sequence $\zeta(t)$, the state's conditional expectation value $\hat{\mathbf{z}}(t)$ and the control input $u(t)$. Figure \ref{fig:3}c shows the quantum trajectory of the particle, which is tracked by the Kalman filter in phase space with the uncertainties in position and momentum given by the diagonal values of the steady-state conditional covariance matrix
$\sigma_z = \sqrt{\hat{{\Sigma}}^\mathrm{ss}_{zz}}= 1.30\, z_{\mathrm{zpf}}$, $\sigma_p = \sqrt{\hat{{\Sigma}}^\mathrm{ss}_{pp}}= 1.35\, p_{\mathrm{zpf}}$ ($p_{\mathrm{zpf}}=\sqrt{\hbar m \Omega_z/2}$: momentum ground-state uncertainty).
To obtain the motional energy of the particle, we evaluate the closed-loop steady-state covariance matrix $\mathbf{\Sigma}^\mathrm{ss}$. For increasing control gain, the mean particle energy $\langle E \rangle=\hbar\Omega_z(\langle n \rangle+1/2)=\hbar\Omega_z\mathrm{tr}(\mathbf{\Sigma}^\mathrm{ss})/2$ ($n$: motional quanta) decreases and quantum ground state cooling ($\langle n \rangle<1$) is achieved for gain levels larger than $2\pi\cdot 40 \mathrm{kHz}$ (Figure \ref{fig:2}d). The estimated occupation values $\langle n \rangle$ agree well with the analytic solution of the LQG problem. We independently confirm these results by Raman sideband thermometry in an out-of-loop heterodyne measurement by mixing the backscattered light with a local oscillator field that is detuned from the trapping field by $\Omega_{\mathrm{het}}=\pm 2\pi\cdot 9.2$MHz (Figure \ref{fig:1}a). This allows us to spectrally resolve the Stokes and anti-Stokes components originating from inelastic scattering off the particle. The scattering rates of these two processes ($\Gamma_{\mathrm{S}}$, $\Gamma_{\mathrm{aS}})$ correspond to the powers detected in the sidebands of the heterodyne measurement. They contain a fundamental asymmetry due to the fact that anti-Stokes scattering, which removes energy from the system, cannot occur from a motional quantum ground state. This is captured by a non-zero difference $\Gamma_{\mathrm{S}}-\Gamma_{\mathrm{aS}}$ of the scattering rates that is independent of the thermal occupation $\langle n \rangle$~\cite{methods} (Figure \ref{fig:3}b). On the other hand, their ratio $\Gamma_{\mathrm{aS}}/\Gamma_{\mathrm{S}} = \langle n \rangle/\left(\langle n \rangle + 1 \right)$ provides us a direct, calibration-free measure of $\langle n \rangle$~\cite{SafaviNaeini2012}.
To exclude other sources of asymmetry that may falsify the measurement, we independently characterize and subtract all (potentially non-white) noise sources (e.g., optical phase noise, detector dark noise) and normalize the data to shot noise, thereby taking into account also the frequency-dependent detector response~\cite{methods}. For consistency, we perform all measurements at both positive and negative heterodyne frequencies. For each gain setting, both measurements agree within the statistical error (Figure \ref{fig:3}b-c~\cite{methods}). 
All data points are also in good agreement with the LQG theory. At maximum gain, we measure a maximal averaged asymmetry of $0.35$, corresponding to an occupation of $\langle n \rangle = 0.56 \pm 0.02$. This establishes quantum ground state cooling of a nanoparticle from room temperature by real-time optimal quantum control. 
In the ideal case, the lowest energy can be achieved at infinite feedback gain and is limited by the steady-state conditional covariance to $\langle n \rangle = 0.34$. In our experiment, the cooling performance is limited by the computational resources of the \textsf{Red Pitaya}, restricting the trade-off between the complexity of the model, the accuracy of the fixed-point arithmetic and the sampling frequency of the implementation. In practice, this generates a significant risk of numerical overflow when the control output is increased above $g_{\mathrm{fb}}=2\pi\cdot200\mathrm{kHz}$.

\section*{Discussion and outlook}

We have demonstrated real-time optimal quantum control of a levitated nanoparticle. Our experiment combines two features: First, using a near Heisenberg-limited confocal measurement scheme, we realize -- at room temperature -- the conditions for which the quantum-mechanical properties of the particle can no longer be neglected~\cite{Braginski1975}. Second, real-time implementation of both a Kalman filter and a linear quadratic regulator (LQR) provides the required algorithms for optimal state estimation and control. As a result, we achieve feedback cooling to the motional quantum ground state ($\langle n \rangle = 0.56 \pm 0.02$) in a room temperature environment.
An immediate application is mechanical sensing of weak stationary~\cite{Ranjit2016,Monteiro2020fs,Moore2021} or transient~\cite{Monteiro2020dm,Carney2020, Moore2021} forces. While neither real-time optimal filtering or feedback cooling improves the signal-to-noise ratio~\cite{Moore2021}, our real-time state estimation can discriminate momentum kicks to the particle
as small as $\Delta p = \sqrt{\sigma^2_p+p^2_{\mathrm{zpf}}} = 1.2\sqrt{\hbar m \Omega_z} = 1.6\times 10^{-23}\,\mathrm{kg\,m/s}$ ($29\ \mathrm{keV/c}$), only a factor 1.2 away from the fundamental quantum limit for continuous  sensing~\cite{Carney2020}. 
This is comparable to the momentum imparted by the inelastic collision with a hydrogen molecule travelling at about $800\,~\mathrm{m/s}$, and smaller than the momentum (in a single dimension) of almost 10\% of the gas molecules at room temperature.
Interestingly, this sensitivity is only a factor of 60 above the latest bounds in the search for gravitationally interacting particle-like candidates for dark matter~\cite{Monteiro2020dm}. In other words, extending our method to particle sizes beyond $1\mu m$ would enable the search for these exotic particles in new parameter regimes. 
From a more general perspective, the ability to drive seemingly classical room temperature objects into genuine quantum states of motion simply by measurement and feedback offers unique possibilities to study quantum phenomena in hitherto unexplored macroscopic parameter regimes~\cite{Leggett2002a,Chen2013}.
Extending our current scheme to a more complex system dynamics may enable the preparation of genuinely non-classical states including squeezed~\cite{Genoni2015} or, in combination with non-linear filtering and anharmonic potential landscapes~\cite{Ralph2018,Rakhubovsky2019b}, even non-Gaussian states of motion.

\paragraph*{Note added.} We recently became aware of a related independent work by Tebbenjohanns \textit{et al.}~\cite{Tebbenjohanns2021}.

\section*{Acknowledgements} We thank José Manuel Leitão for his introduction to optimal control, and Paolo Vezio, Hans Hepach and Tobias Westphal for discussions and their help in the lab. L.~M. thanks Arno Rauschenbeutel for the discussion inspiring the confocal detection scheme. 
This project was supported by the European Research Council (ERC 6 CoG QLev4G), by the ERA-NET programme QuantERA under the Grants QuaSeRT and TheBlinQC (via the EC, the Austrian ministries BMDW and BMBWF and research promotion agency FFG), by the European Union’s Horizon 2020 research and innovation programme under Grant No. 863132 (iQLev), and by the Austrian Science Fund (FWF, START Project TheLO, Y 952-N36). L. M. is supported by the Vienna Doctoral School of Physics (VDS-P) and by the FWF under project W1210 (CoQuS).

\section*{Author contributions} L.~M. designed and built the experiment, P.~R. designed and programmed the filter and controller. L.~M. and C.~B. performed the measurements. L.~M., P.~R. and C.~B. analyzed the data and all authors contributed to writing and editing of the paper. 

\section*{Data Availability} The data that support the plots within this paper and other findings of this study are available from the corresponding author upon reasonable request.

\section*{Competing Interests} The authors declare no competing financial interests.

\let\oldaddcontentsline\addcontentsline
\renewcommand{\addcontentsline}[3]{}
\bibliographystyle{bibliography_nature}
\bibliography{Bibliography}
\let\addcontentsline\oldaddcontentsline


\clearpage
\onecolumngrid

\renewcommand{\thesection}{A\arabic{section}}  
\renewcommand{\thetable}{A\arabic{table}}  
\renewcommand{\thefigure}{A\arabic{figure}} 
\renewcommand{\theequation}{A\arabic{equation}} 
\setcounter{figure}{0}
\setcounter{equation}{0}
\setcounter{section}{0}
\setcounter{table}{0}
\hypersetup{colorlinks,linkcolor={red!50!black},citecolor={blue!50!black},urlcolor={blue!80!black}}

\begin{center}
\large\textbf{\uppercase{Appendix}}
\end{center}

\vspace{3,5cm}
\tableofcontents
\vspace{3,5cm}
\section{The complete experimental setup}
\label{sec:setup}

We include the complete experimental details of the experiment. The core of the experiment is the optical tweezer: a microscope objective of NA = 0.95 to tightly focusing $\sim 300\,\mathrm{mW}$ of light at $\lambda = 1064\,\mathrm{nm}$ ($\omega_0 = c2\pi/\lambda$: optical frequency) in ultra-high vacuum. Before reaching the optical trap, part of the light is diverted to a couple of acousto-optic modulators (AOMs) oriented to scatter in positive and negative first order. In order to avoid slow intensity drifts due to interference between the parametric feedback and the optical tweezer, the parametric feedback cooling is implemented with light at optical frequencies of $\omega_{\mathrm{pfb}}=\omega_0+2\pi\cdot 205\,\mathrm{MHz}$ while the light shifted by the second AOM at a frequency of $\omega_{\mathrm{het}}=\omega_0\pm 2\pi\cdot 9.2\,\mathrm{MHz}$ is used as local oscillator for the heterodyne measurement.  The polarization in the tweezer is controlled by a half and a quarter waveplate (HWP,QWP). This allows us to excite the rotational degree of freedom and its precession about the $z$-axis in ultra-high vacuum  to frequencies above 100 MHz by briefly applying an optical torque to the levitated particle, avoiding disturbance at the frequencies of interest. The back scattered light is selected by a Faraday rotator (FR) and a polarizing beam-splitter (PBS) and routed to the confocal fiber filtering. Here a lens ($f_3$) focuses light into a single-mode fiber (green). A variable ratio coupler (VRC) is used to split the light between homodyne and heterodyne detection. The use of these tunable VRCs, also in the actual interferometric measurement, allows us to balance the splitting ratio with a precision below 0.5\%.
The slow phase drift of the homodyne signal is stabilized by use of a low-pass filter (LP) and PID controller driving a fiber stretcher constituted of a bare fiber wrapped around a cylindrical piezo. The signal is then directed to the Red-Pitaya (RP) board which calculates the state estimates and a calibrated control signal. The control signal is applied to the holder of a collection lens which serves as electrode and is placed in front of the tweezer objective which is grounded~\cite{Frimmer2017}. The homodyne and heterodyne measurement sequences as well as  the state estimates and control signals are recorded simultaneously. After the tweezer, light is collected by a lens and used for 3D forward split-detection (BS: beam splitter). This low quality measurement serves to implement the parametric feedback of all 3 modes: a phase lock loop (PLL) allows to track the phase of each mode and stabilize its motion by modulating the optical spring at twice the mechanical frequency via an electro-optic modulator (EOM) and overlayed with the tweezer light by a PBS.
During the experiment, the parametric feedback for the $z$-mode is switched off. A green laser is shined from the side onto the particle for imaging of the dipole scattering through a dichroic mirror (DM) onto a CCD sensor.
\begin{figure}[!h]
\includegraphics[scale=1]{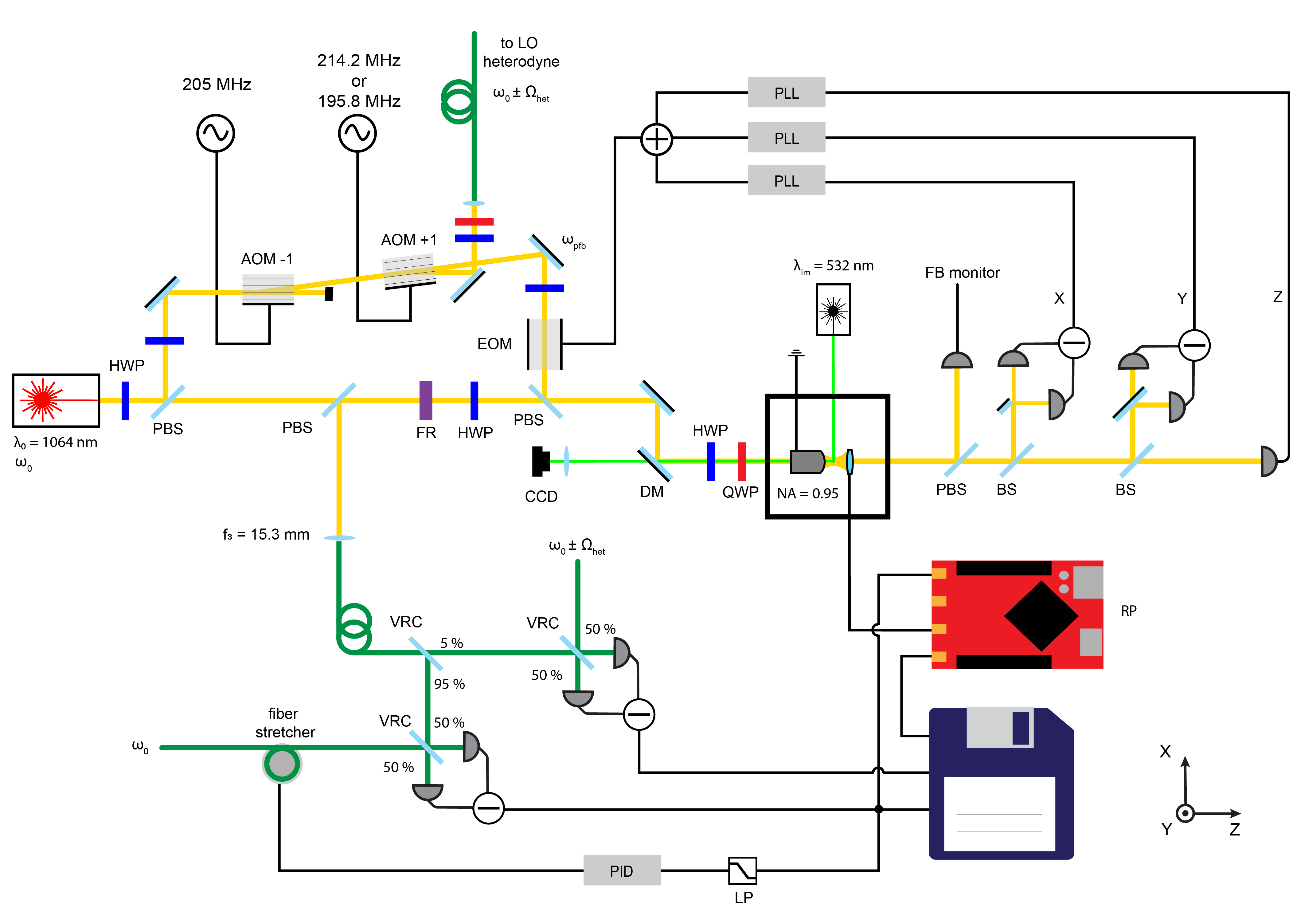}
\caption{\textbf{The experimental setup}.}
\label{fig:setupcomplete}
\end{figure}

\section{Imprecision and backaction noise in an optical tweezer}
\label{sec:theorysql}
We describe the effects of quantum noise in a measurement process following the description by Clerk \textit{et al.}~\cite{Clerk2010} for a flat mirror moving in one dimension. We then extend this to the geometry of an optically levitated particle, along the lines of the analysis showed by Seberson and Robischeaux~\cite{Seberson2019}. A full quantum description of the open quantum system in terms of the quantum Langevin equations and input-output formalism will be derived in Section~\ref{sec:qlangevin}.

When performing a phase measurement of light in a coherent state (displaced vacuum) the phase and photon number uncertainty is governed by Poissonian statistics: these uncertainties are respectively $\Delta \varphi  = 1/(2\sqrt{N})$ and $\Delta N  = \sqrt{N}$, where $N$ is the measured number of photons during the time $t$. The product of these uncertainties satisfies the relation $\Delta N\Delta \varphi =1/2$~\cite{Loudon}. 
In the context of continuous measurements of stationary processes it is useful to reformulate these quantities in terms of a noise power spectral densitiy. This is defined, for a variable $X$, as the Fourier transform of its autocorrelation:
\begin{equation}
    S_{XX}(\Omega) = \int\limits_{-\infty}^{+\infty} e^{-\ii\Omega t}  \left\langle X(0)X(t)  \right\rangle\,dt
\end{equation}
Measuring a continuous flux of photons of average $\bar{\dot{N}}$, we can now define $S_{\varphi\varphi} = (\Delta \varphi)^2/t = 1/(4 \bar{\dot{N}})$ and $S_{\dot{N}\dot{N}} = (\Delta N)^2/t = \bar{\dot{N}}$. Again, we have the uncertainty relation:
\begin{equation}
\sqrt{S_{\varphi\varphi}S_{\dot{N}\dot{N}}} = 1/2
\end{equation}

\subsection{Measuring the displacement of a flat mirror}
As a first example of optical measurement, we study the one dimensional case of a photon bouncing off a mirror, measuring its displacement $x$. The phase shift gained by each photon is two times the phase shift acquired in $x$ distance: $\varphi = 2kx$. The momentum transferred to the mirror by elastic scattering is twice the photon momentum $p=2\hbar k$. These lead to spectral density definitions for imprecision of position measurement and random backaction force-noise: $S_{xx}^\mathrm{I}=S_{\varphi\varphi}/(4k^2)$ and $S_{FF}^\mathrm{ba}=4\hbar^2k^2 S_{\dot{N}\dot{N}}$~\cite{Clerk2010}. The uncertainty relation becomes:
\begin{equation}
    \sqrt{S_{xx}^\mathrm{I} S_{FF}^\mathrm{ba}}=\hbar/2
\label{eq:heisnberg}
\end{equation}

\subsection{The case of a levitated particle}

The case of a levitated particle is qualitatively equivalent. The main difference to consider is that the direction of incoming and scattered photons is not necessarily parallel to the direction of the particle's motion that we are interested in measuring. The total optomechanical interaction is distributed to the different degrees of freedom ($x$, $y$, $z$), reducing the average coupling to each mode.
We follow the same steps and notation as described by Seberson and Robicheaux~\cite{Seberson2019} to derive the imprecision noise and measurement backaction for our system. 
A photon of initial wave-vector $\roarrow{k}_\mathrm{i} = k\hat{k}_\mathrm{i} = (k_{\mathrm{i}x},k_{\mathrm{i}y},k_{\mathrm{i}z})= k(\hat{k}_{\mathrm{i}x},\hat{k}_{\mathrm{i}y},\hat{k}_{\mathrm{i}z})$ scatters elastically off a particle at position $\roarrow{r}$, initial velocity $\roarrow{v}_\mathrm{i}$ and mass $m$. The phase shift acquired by the photon and the final velocity of the particle are given by:
\begin{equation}
\varphi = \roarrow{k}_\mathrm{i}\roarrow{r}-\roarrow{k}_\mathrm{f}\roarrow{r}
\quad \mathrm{and}\quad
\roarrow{v}_\mathrm{f} = \roarrow{v}_\mathrm{i}+ \frac{\hbar}{m}\left(\roarrow{k}_\mathrm{i} - \roarrow{k}_\mathrm{f} \right), 
\label{eq:phaseandmomentum}
\end{equation}
where $ \roarrow{k}_\mathrm{f} $ is the final wave-vector. The squared phase shift resulting from a displacement along the direction $j = (x,y,x)$ is:
\begin{equation}
 \varphi_j^2 =
\left(k_{\mathrm{i}j} r_j - k_{\mathrm{f}j} r_j \right)^2 =
 k^2 r_j^2 \left(  \hat{k}_{\mathrm{i}j}-\hat{k}_{\mathrm{f}j} \right)^2,
\label{eq:paseshift2}
\end{equation}
and similarly the square momentum exchanged with the particle's mode $j$ is:
\begin{equation}
 p_j^2 =
 \left(m v_{\mathrm{f}j} - m v_{\mathrm{i}j} \right)^2  = 
 \hbar^2 k^2 \left( \hat{k}_{\mathrm{i}j}-\hat{k}_{\mathrm{f}j} \right)^2,
\label{eq:momentum2}
\end{equation}
where we have used the fact that $\langle v_j \rangle = 0$ for harmonic motion.
\begin{figure}[!h]
\includegraphics[scale=1]{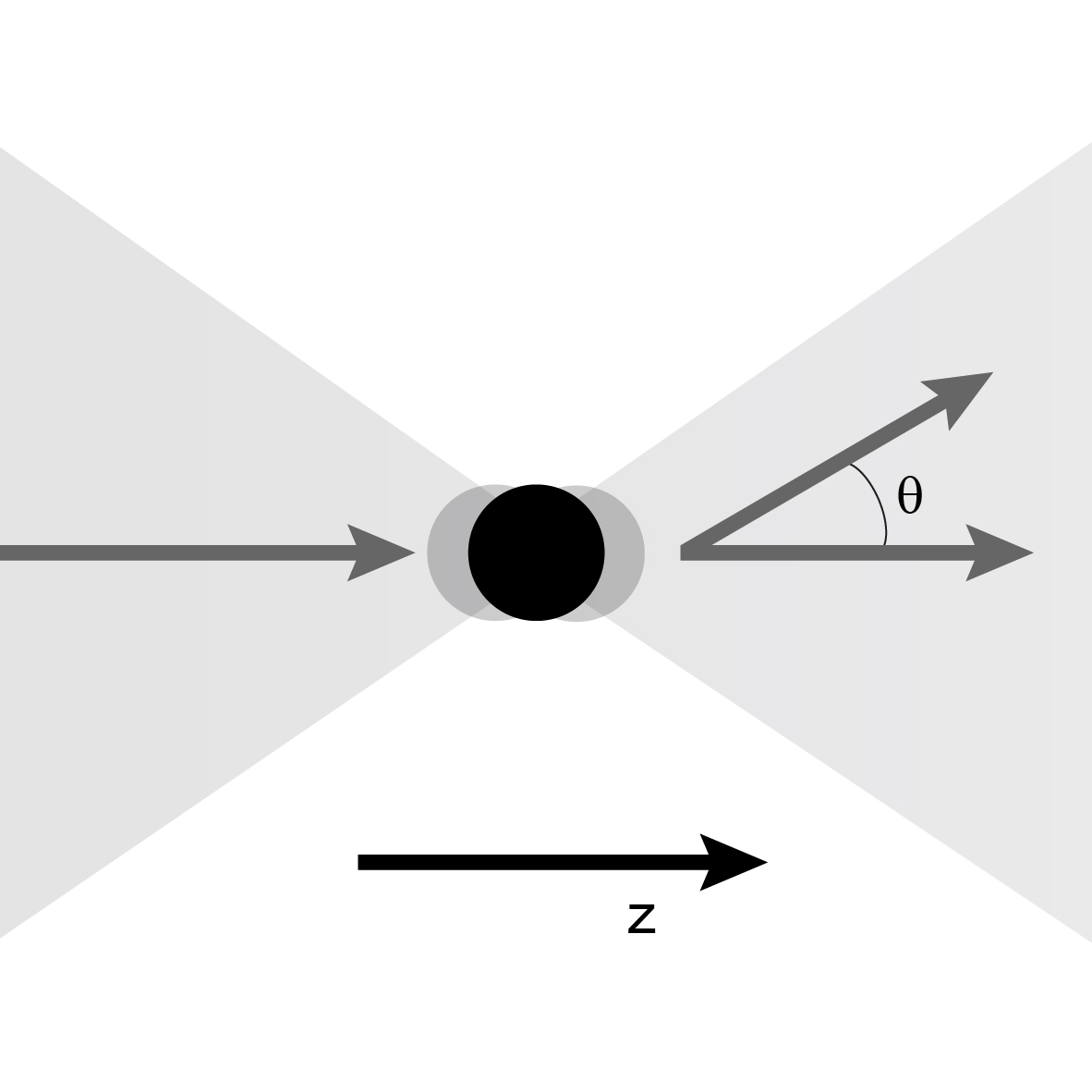}
\caption{\textbf{Scattering angle}. We define the scattering angle $\theta$ as the angle between the $z$ axis and the scattering direction.}
\label{fig:geometry}
\end{figure}
\noindent
As the phase and momentum depend on the incidence and scattering directions, in order to compute second moments of momentum and phase fluctuations we have to consider the scattering probability distribution defined for a dipole emitter. The probability of a photon emitted by a dipole being scattered in direction $\hat{k}_\mathrm{f}$ is~\cite{Seberson2019, Tebbenjohanns2019_detection}:
\begin{equation}
P(\hat{k}_\mathrm{f})=\frac{3}{8\pi}(\cos^2\theta \cos^2\phi +\sin^2\phi),    
\end{equation}
where the spherical coordinate system is defined such that the scattered photon has the direction $\hat{k}_\mathrm{f} = (\sin \theta \cos\phi,\  \sin \theta \sin\phi,$ $\cos\theta)$. Note that $\int\limits_{4\pi }P(\hat{k}_\mathrm{f}) d\Omega = 1$.
For each direction of motion, the sqaure optical phase shift and momentum exchange, averaged over the scattering probability distribution is then given by:
\begin{equation}
   \langle  \varphi ^2_j\rangle = \int\limits_{4\pi} P(\hat{k}_\mathrm{f})  \varphi_j^2 d\Omega
   \quad \mathrm{and}\quad
   \langle  p ^2_j\rangle = \int\limits_{4\pi} P(\hat{k}_\mathrm{f})  p_j^2 d\Omega.
\label{eq:integral}
\end{equation}
We consider the case for $j=z$, which is the direction of interest of this paper. The other directions follow trivially and are discussed in \cite{Seberson2019, Tebbenjohanns2019_detection}. If considering an incident plane wave, the incidence and scattering wave vectors are defined as $\hat{k}_{\mathrm{i}z} = 1$ and $\hat{k}_{\mathrm{f}z} = \cos\theta$, respectively. As we are dealing with a tightly focused beam we have to modify the value of the initial wave vector to be $\hat{k}_{\mathrm{i}z}= A \leq 1$, where A is a geometrical factor arising from the Gouy phase shift in the focal field which depends on the trapping NA and can be computed following~\cite{Tebbenjohanns2019_detection}. In our case $A=0.71$. Inserting \eqref{eq:paseshift2} and \eqref{eq:momentum2} into \eqref{eq:integral}, and projecting on $z$,
\begin{equation}
    \langle  \varphi ^2_z\rangle = k^2 z^2 \int\limits_{4\pi} P(\hat{k}_\mathrm{f}) (A-\cos\theta)^2 d\Omega
    \quad \mathrm{and} \quad
    \langle p ^2_z\rangle = \hbar^2 k^2 \int\limits_{4\pi} P(\hat{k}_\mathrm{f}) (A-\cos\theta)^2 d\Omega
\label{eq:imprecisionbackaction}
\end{equation}
The mean square phase shift and square momentum exchange along $z$ become: 
\begin{equation}
     \langle  \varphi ^2_z\rangle = \left(A^2+\frac{2}{5}\right)k^2 z^2
     \quad\mathrm{and}\quad
      \langle  p ^2_z\rangle = \left(A^2+\frac{2}{5}\right)\hbar^2 k^2
\end{equation}
As for the one dimensional case we can now express the interaction in terms of spectral densities for position imprecision and force noise, extending the averaging to the time domain:
\begin{equation}
     S_{zz}^\mathrm{I} = \frac{ S_{\varphi\varphi}}{\left( A^2 + \frac{2}{5}\right) k^2} 
     \quad\mathrm{and}\quad
      S_{FF}^\mathrm{ba}=\left(A^2+\frac{2}{5} \right)\hbar^2k^2 S_{\dot{N}\dot{N}}
\end{equation}
and in terms of optical scattered power, $P_\mathrm{scatt} = \hbar \omega \bar{\dot{N}}= \hbar c k \bar{\dot{N}}$:
\begin{equation}
     S_{zz}^\mathrm{I} = \frac{\hbar c}{\left( A^2 + \frac{2}{5}\right) 4k P_\mathrm{scatt}} 
     \quad\mathrm{and}\quad
      S_{FF}^\mathrm{ba}=\left(A^2+\frac{2}{5} \right)\frac{\hbar k P_\mathrm{scatt}}{c},
      \label{eq:SD_ForceImprecision}
\end{equation}
which also fulfills the Heisenberg uncertainty relation \eqref{eq:heisnberg}. 

As we measure real signals, it is useful to consider the one-sided power spectral density, defined for a real signal $X$, at positive frequencies as: 
\begin{equation}
  S_X(\Omega \geq 0 ) = \left(S_{XX}(\Omega) + S_{XX}(-\Omega)\right)
\end{equation}
where the variance of the signal $X$ is:
\begin{equation}
    \left\langle X^2 \right\rangle = \frac{1}{2\pi}\int\limits_{-\infty}^{+\infty}S_{XX}(\Omega)d\Omega = \frac{1}{2\pi}\int\limits_{0}^{+\infty}S_X(\Omega)d\Omega
\end{equation}
which for a real white process simply reduces to $S_{X} = 2S_{XX}$.
In terms of single-sided power spectral densities, the uncertainty relation becomes:
\begin{equation}
    \sqrt{S_z^\mathrm{I}S_F^\mathrm{ba}}=\hbar
\label{eq:heisnbergsym}
\end{equation}
In real experiments the backaction-imprecision product is degraded by losses. On the one hand, there are losses of information in the detection channel $\eta_d$. They increase the imprecision noise while leaving the backaction force noise unaltered; The detected imprecision noise becomes $S_z^\mathrm{imp} = S_z^\mathrm{I}/\eta_\mathrm{d}$. On the other hand, there are losses of information by interactions with the environment $\eta_\mathrm{e}$. Environmental force noise contributions include scattering of gas molecules, feedback noise, black-body radiation; All having the effect of exchanging momentum with the system, without contributing to the measurement. The total force noise becomes $S_F^\mathrm{tot} =\sum_i S_F^{i} = S_F^\mathrm{ba}/\eta_e$ and the imprecision-backaction product can be written as:
\begin{equation}
    \sqrt{S_z^\mathrm{imp} S_F^\mathrm{tot}} = \frac{\hbar}{\sqrt{\eta}} \geq \hbar
\label{eq:heisnbergsymloss}
\end{equation}
where $\eta = \eta_\mathrm{d}\eta_\mathrm{e}$ considers information losses in the detection and into the environment. In the following sections (\ref{sec:losses}, \ref{sec:forcenoise}), we will analyze losses in the detection channel and discuss additional force noise contributions to the backaction term. 

\subsection{The standard quantum limit for the harmonic oscillator}

\begin{figure}[!htb]
\includegraphics[width=13cm]{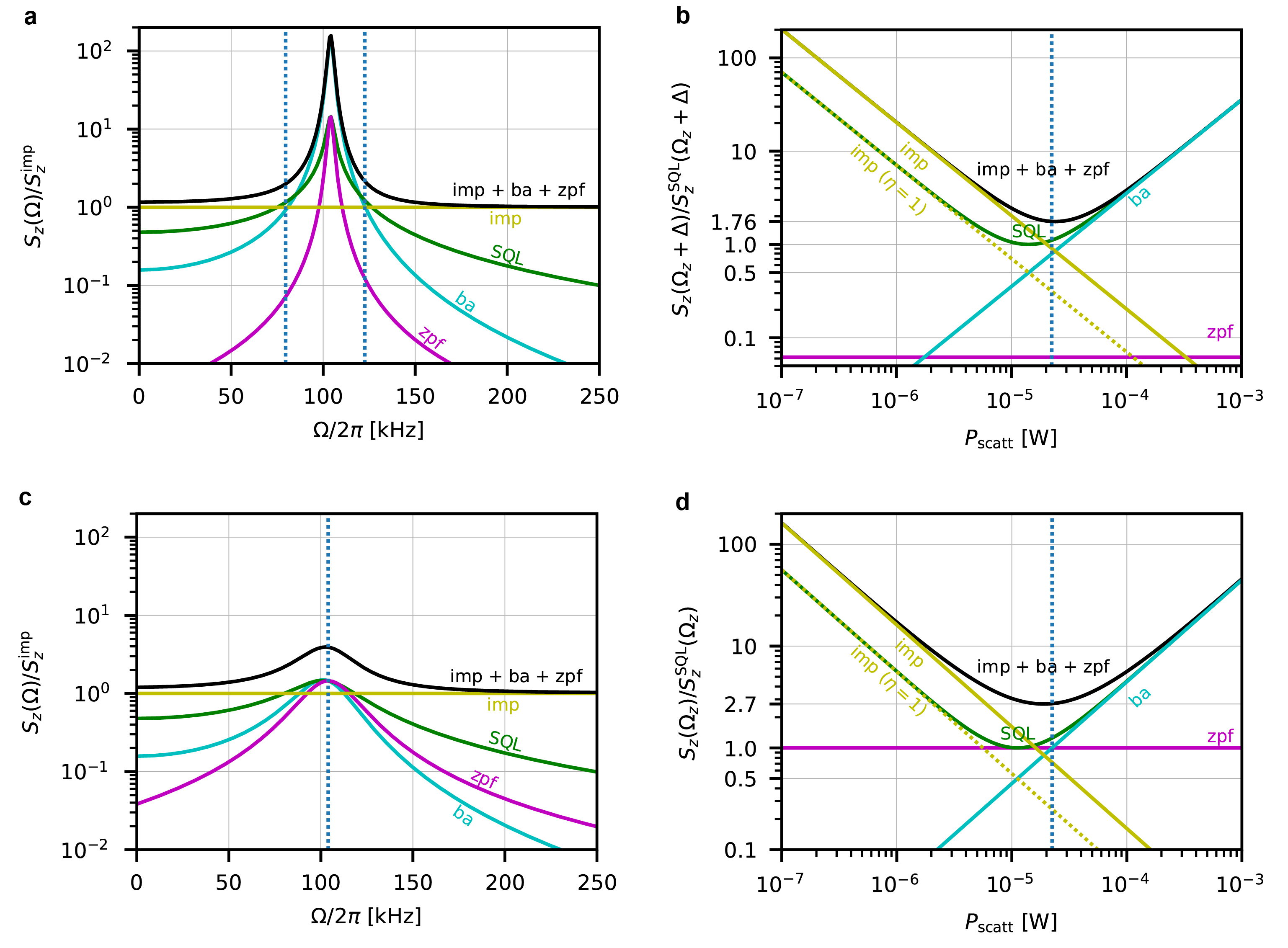}
\caption{\textbf{The standard quantum limit}. \textbf{a}, Contribution to the measured power spectral density of imprecision (imp), backaction (ba), and zero point fluctuation (zpf), compared to the SQL as a function of frequency, in a regime of weak cooling ($\Gamma_{\mathrm{fb}}=\Gamma_{\mathrm{ba}}/5$). \textbf{b}, Contribution to the total noise, evaluated at $\Omega_z \pm \Delta$ (verical dotted line in \textbf{a}), as a function of the scattered power (measurement strength). In the case of weak cooling, the contributions of backaction and zero point fluctuation are concentrated on resonance, allowing perfect balancing of imprecision and backaction when $\Delta \approx 2\pi\times 22$~kHz, and resulting in a total added noise that is only a factor 1.76 from the SQL. \textbf{c}, Contributions to the measured power spectral density in a regime of strong cooling ($\Gamma_{\mathrm{fb}}=2\Gamma_{\mathrm{ba}}$). In this case, the contributions of backaction and the zero point fluctuations are broadened in frequency, allowing on resonance (vertical dotted line), a suppression of the added noise to a factor of 2.7 from SQL (in \textbf{d} as a function of the scattered power). Note that in the case of optical tweezers, \textbf{b} and \textbf{d} do not represent a complete set of experimentally available conditions, and are only valid at a fixed scattered power (vertical dotted lines). A variation of this would necessarily come along with a change in the mechanical frequency, and a redefinition of the system parameters. This representation is however useful to understand the operating conditions of the system with respect to the SQL.}
\label{fig:sql}
\end{figure}

The response of a system to external forces is given by its mechanical susceptibility, defined, for a harmonic oscillator, as: $\chi_\mathrm{m}(\Omega) = [m(\Omega_z^2 - \Omega^2 + \ii\gamma\Omega)]^{-1}$ ($m$ the mass of the particle, $\Omega_z$: the mechanical resonance frequency, $\gamma$: the total damping of the system). The relation between imprecision and backaction (equation \eqref{eq:heisnbergsym}) defines a minimal added noise to the measured displacement spectrum that is known as the standard quantum limit. This limit is achieved, at a given frequency, when the strength of measurement is such that the contributions of imprecision and response to backaction are equal~\cite{Mason2019}:
\begin{equation}
S_{z}^\mathrm{SQL}(\Omega) = \min \{ S^\mathrm{I}_z +S_{F}^\mathrm{ba}\lvert \chi_\mathrm{m}(\Omega)\rvert^2\} = 2 \hbar |\chi_\mathrm{m}(\Omega)|
\end{equation}
In a real measurement one has to consider not only losses in the detection and environmental force noise contributions, but also the oscillator's quantum fluctuations of position $z_\mathrm{zpf} = \sqrt{\hbar/(2m\Omega_z)}$, resulting in a ground state displacement spectrum~\cite{Clerk2010}:
\begin{equation}
S_{z}^\mathrm{zpf}(\Omega)  = z_\mathrm{zpf}^2\frac{\gamma}{\left(\Omega-\Omega_z\right)^2 +\left(\gamma/2\right)^2}.
\end{equation}
The particle motional spectrum is:
\begin{equation}
    S_z(\Omega) =  S_{F}^\mathrm{tot}\lvert \chi_\mathrm{m}(\Omega)\rvert^2+ S_z^\mathrm{zpf}(\Omega)
    \label{eq:motionpsd}
\end{equation}
and the total measured displacement noise then becomes: 
\begin{equation}
S_\zeta(\Omega) = S^\mathrm{imp}_z +  S_{F}^\mathrm{tot}\lvert \chi_\mathrm{m}(\Omega)\rvert^2+ S_z^\mathrm{zpf}(\Omega)
\end{equation}
where $\zeta = z + \nu$ is the sum of the actual motion of the particle together the position equivalent measurement noise. It is evident that, in the case of weak damping, backaction and quantum fluctuations have a large contribution to the total noise on resonance, and the added noise is much larger than the SQL (Figure \ref{fig:sql}a, \ref{fig:sql}c). Off resonance however it is possible to find frequencies where the noise is closest to the SQL (Figure \ref{fig:sql}b). Up to a certain degree it is also possible to suppress the backaction contribution on resonance, and redistribute the quantum zero point fluctuation noise contribution to a larger frequency band. This is done by feedback cooling which increases damping and modifies the mechanical susceptibility (Figure \ref{fig:sql}c, \ref{fig:sql}d and Section \ref{sec:forcenoise}).
In our system, with an information efficiency of $\eta = 0.34$, we distinguish 2 regimes: the weakly cooled regime where we achieve (off resonance) a displacement noise of 1.76 times the SQL, and a strongly cooled regime, where by strongly suppressing backaction we are able to achieve (on resonance) a displacement noise that is 2.7 times the SQL. Note that for the resonant case, even at zero temperature the contribution of the zero point fluctuations limits the displacement noise to 2 times the SQL. These results show an improvement of more than one order of magnitude for a mechanical system at room temperature~\cite{Abbott2009, Bushev2013, Schilling2016,Tebbenjohanns2020, Kamba2020}.
\begin{figure}[!htb]
\includegraphics[width=14cm]{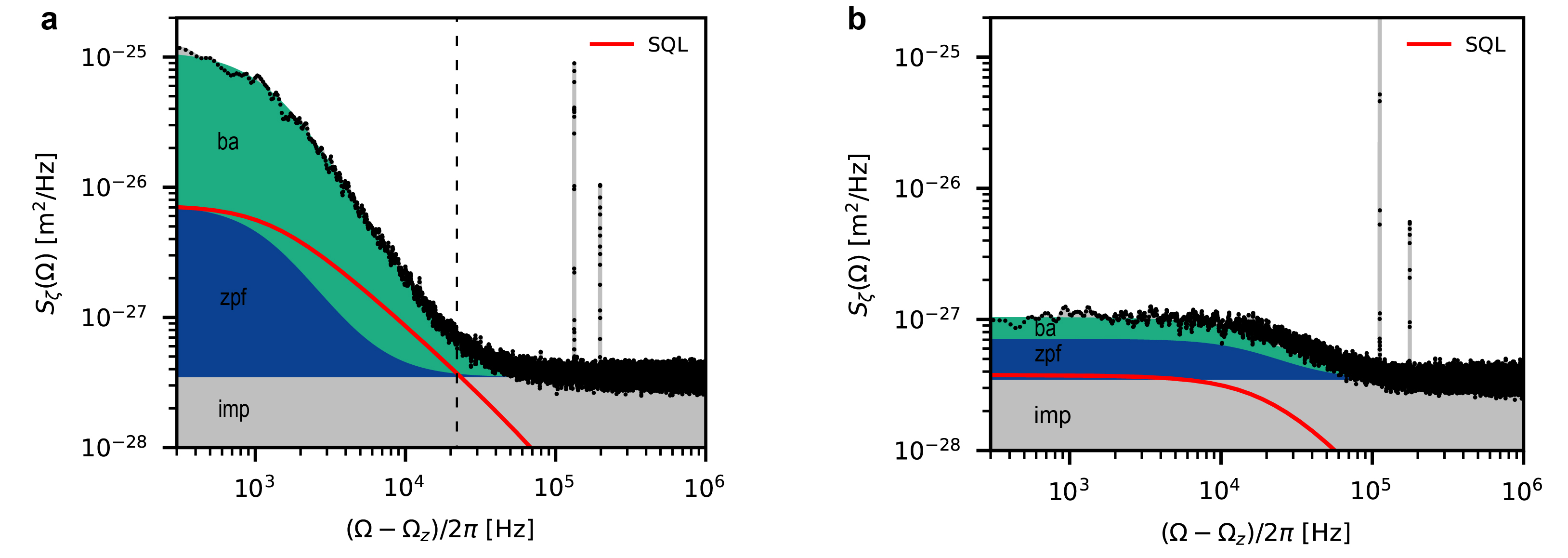}
\caption{\textbf{The measured noise}. Measured displacement power spectral density (black) showing the contributions by imprecision (imp, gray), backaction (ba, green), and the zero point fluctuations (zpf, blue), compared to the SQL (red). \textbf{a} A feedback gain of $g_{\mathrm{fb}}/2\pi = 2\, \mathrm{kHz}$ results in an occupation of $n = 8.3\pm 0.09$. The almost perfect balancing of imprecision and backaction at $22\,\mathrm{kHz}$ above resonance (vertical dashed line) results in a measurement that is only a factor 1.76 from the ideal SQL. \textbf{b} In the case of strong cooling ($g_{\mathrm{fb}}/2\pi = 110\, \mathrm{kHz}$), and occupation of $n = 0.71\pm 0.09$, we achieve a total added noise on resonance that is a factor 2.7 higher than the SQL.}
\label{fig:sql2}
\end{figure}

\subsection{Measurement and decoherence rates}

The resolution of a noisy measurement increases with measurement time. A quantum limited measurement however necessarily disturbs the system, limiting the time for which one can measure a quantum state before it is completely destroyed by the measurement itself~\cite{Clerk2003}. We introduce rates of measurement and decoherence to quantify these processes. We define the measurement rate as the rate at which our measurement is able to resolve a displacement equivalent to the zero point motion of the particle ($z_\mathrm{zpf})$:
\begin{equation}
    \Gamma_\mathrm{meas} = \frac{z^2_\mathrm{zpf}}{4S^{\mathrm{imp}}_{zz}} = \frac{z^2_{\mathrm{zpf}}}{2S^{\mathrm{imp}}_z} =\eta_\mathrm{d}\frac{z^2_{\mathrm{zpf}}}{2S^{\mathrm{I}}_z} 
\end{equation}
Similarly, the decoherence rate, defined as the rate of energy quanta delivered to the oscillator by the measurement process, is 
\begin{equation}
    \Gamma_\mathrm{ba} = \frac{S^{\mathrm{ba}}_{FF}}{4p^2_{\mathrm{zpf}}} = \frac{S^{\mathrm{ba}}_F}{8p^2_{\mathrm{zpf}}}
\end{equation}
where $p_\mathrm{zpf} = \sqrt{\hbar m\Omega_z/2}$ momentum ground-state uncertainty. With the help of equation \eqref{eq:heisnbergsym}, we can compute the ratio of measurement rate and backaction-induced decoherence rate:
\begin{equation}
    \frac{\Gamma_{\mathrm{meas}}}{\Gamma_{\mathrm{ba}}} = \eta_\mathrm{d} \leq 1.
    \label{eq:deteff}
\end{equation}
Decoherence in the system, however, does not only originate from the measurement process, but also from other environmental interactions. We define the rate of decoherence induced by the environment (commonly thermal) as $\Gamma_{\mathrm{th}}$. The strength of a measurement with respect to other environmental interactions is known as the measurement quantum cooperativity: $C_q = \Gamma_{\mathrm{ba}} /\Gamma_{\mathrm{th}}$. Finally, using equation \eqref{eq:deteff} it is possible to define the measurement information efficiency, which summarizes the quality of a measurement process:
\begin{equation}
    \eta = \frac{ \Gamma_{\mathrm{meas}} }{ \Gamma_{\mathrm{ba}} + \Gamma_{\mathrm{th}} } = \eta_\mathrm{d} \frac{ \Gamma_{\mathrm{ba}} }{ \Gamma_{\mathrm{ba}} + \Gamma_{\mathrm{th}}} = \eta_\mathrm{d} \left(1+\frac{1}{C_q}\right)^{-1}  = \eta_\mathrm{d}\eta_\mathrm{e}
\end{equation}

\subsection{Noise equivalent occupation}

When monitoring the position of a harmonic oscillator, often the quantities of imprecision and force noise are considered in units of energy quanta. We can assign an apparent thermal occupation to the imprecision noise~\cite{Wilson2015,Suhdir2017,Rossi2018}:
\begin{equation}
    n_\mathrm{imp} = \frac{S_{z}^\mathrm{imp}}{2S_{z}^\mathrm{zpf}(\Omega_z)} = S_{z}^\mathrm{imp}\frac{\gamma}{8 z^2_\mathrm{zpf}} 
    \label{eq:nimp}
\end{equation}
On the other hand we can assign an occupancy to the bath associated with the force noise driving the oscillator. Assuming energy equipartition this is:
\begin{equation}
    n_\mathrm{tot} = \frac{1}{2\pi}\int\limits_0^\infty  \frac{S_{F}^\mathrm{tot}\lvert \chi_\mathrm{m}(\Omega)\rvert^2}{2 z^2_\mathrm{zpf}} d\Omega = \frac{S_{F}^\mathrm{tot}}{8 p^2_\mathrm{zpf}\gamma}
    \label{eq:ntot}
\end{equation}
where the last identity in equation \eqref{eq:ntot} is only valid in the case of a white force noise. 
The effect of backaction associated to any quantum measurement process seemingly would prohibit any kind of quantum control. However, the effects of this noise are directly captured by the measurement, and can be counteracted by feedback control schemes.
We can then write the minimal achievable occupancy in presence of an ideal feedback as~\cite{Wilson2015}:
\begin{equation}
n_\mathrm{min} = 2 \sqrt{n_\mathrm{imp} n_\mathrm{tot}} -\frac{1}{2}
\label{eq:nmin}
\end{equation}
Note that equation \eqref{eq:nmin} is an asymptotic value, requiring an experimentally impractical infinite bandwidth feedback (see also Section \ref{sec:kalman}). Given the parameters in our system we estimate $n_\mathrm{min} = 0.34$.
\section{Losses of information and photons}
\label{sec:losses}
As we have seen in the previous section \ref{sec:theorysql}  information is not uniformly distributed across the dipole scattered light. Whenever there are spatially dependent losses, there are mismatches between the loss of photons and the actual loss of information. In other words, there are losses and the are losses. We will refer to efficiency that is complementary to information loss with $\eta$ and efficiency that is complementary to photon loss with $\eta^*$.

\subsection{Microscope collection}

The collection efficiency by the microscope objective of dipole scattered photons is:
\begin{equation}
    \eta^*_\mathrm{d,c} = \frac{\int\limits_{\Omega_\mathrm{coll}} P(\hat{k}_\mathrm{f}) d\Omega}{\int\limits_{4\pi} P(\hat{k}_\mathrm{f}) d\Omega}, 
\end{equation}
which results in a photon collection efficiency of $\eta^*_\mathrm{d,c} = 0.375$.
On the other hand, the information collection efficiency by the microscope objective is the ratio of the imprecision noise calculated for a limited collection angle $\Omega_\mathrm{coll}$ over the ideal imprecision noise defined in equation~\eqref{eq:imprecisionbackaction}
\begin{equation}
    \eta_\mathrm{d,c} = \frac{\int\limits_{\Omega_\mathrm{coll}} P(\hat{k}_\mathrm{f}) (A-\cos\theta)^2 d\Omega}{\int\limits_{4\pi} P(\hat{k}_\mathrm{f}) (A-\cos\theta)^2 d\Omega}.
    \label{eq:etacoll}
\end{equation}
With an NA of 0.95 this leads to an information collection efficiency of $\eta_\mathrm{d,c} = 0.84$.

\subsection{Confocal mode-matching}

After being collected by the microscope objective, light needs to be matched to the local oscillator. We implement a fiber based confocal dipole detection~\cite{Vamivakas2007}. This has two advantages: first it allows easy and efficient mode matching of the dipole scattered light to the local oscillator, second, confocal filtering by the fiber allows to suppress stray reflections in the trapping-detection path. Following the description by Vamivakas \textit{et al.}~\cite{Vamivakas2007} we compute the mode overlap between the electric dipole  far field $E_\mathrm{dip}$ imaged at the fiber boundary and the fiber mode profile $E_\mathrm{fm}$ in cylindrical coordinates as a function of magnification $M = f_3/f_1$. Here $f_1$ and $f_3$ are the focal lengths of the objective lens and the imaging lens respectively. The mode overlap efficiency is defined as:
\begin{equation}
    \eta^*_\mathrm{d,m}( M) =\frac{\lvert \int \vec{E}_\mathrm{dip}^*(\vec{r_3}) \vec{E}^x_\mathrm{fm}(\vec{r}_3) dA_3 \rvert^2 }{ \int\lvert \vec{E}_\mathrm{dip}^x(\vec{r_3}) \rvert^2 dA_3 \int\lvert \vec\vec{E}^x_\mathrm{fm}(\vec{r}_3) \rvert^2 dA_3},
    \label{Eq:Overlap}
\end{equation}
where the dipole is oriented along $\hat{x}$ with its origin in the focal point of a 0.95 NA microscope objective and the fiber mode superscript $x$ indicates the x polarized solution. We integrate the overlap of dipole image and fiber mode over the fiber tip surface $dA_3$ at the focal position.
A maximal collection efficiency of 0.76 can be achieved with a magnification of $f_3/f_1 \approx 7.7$.
In our case a magnification of $M = 8.5$ leads to a mode matching efficiency of $\eta^*_\mathrm{d,m} = 0.75$. We manage to couple up to $\eta^*_\mathrm{d,m} = 0.71$.

\begin{figure*}[!htb]
\includegraphics[scale=1]{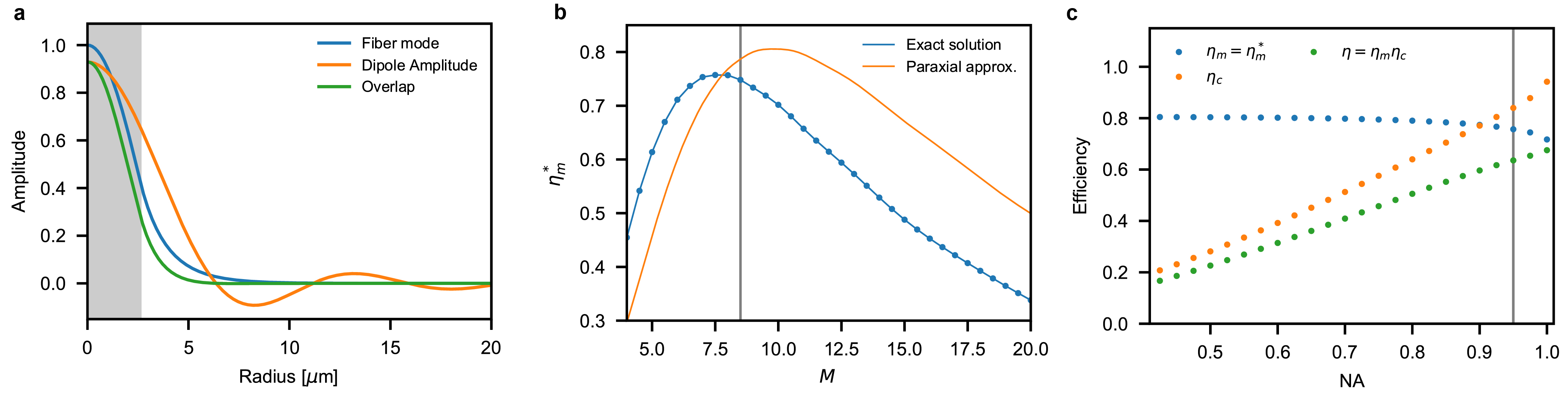}
\caption{\textbf{Fiber-dipole mode overlap}. \textbf{a}, Numerical calculation of the dipole mode (orange) at a fixed angle imaged at the fiber interface by our confocal microscope system, fiber mode (blue), and their overlap (green) as a function of the distance to the center of the fiber. The gray shaded area represents the fiber core. \textbf{b}, Overlap efficiency as a function of magnification of the optical system. The gray vertical line shows our operating point, not far from the optimal value. \textbf{c}, Information collection efficiency by the microscope objective (orange dots), maximum fiber mode matching (blue dots)  and the product of the two (green dots) as a function of the objective NA. The gray line is our operating point.}
\label{Fig:Overlap}
\end{figure*}

For comparison, we also calculate the overlap integral for the dipole image in paraxial approximation, where the collection angle $\theta \rightarrow 0$. The x component of the dipole image becomes:
\begin{equation}
    E^x_\mathrm{dip}(\rho_3,M) = \theta_1 \frac{M}{k_3 \rho_3} J_1(k_3 \rho_3 \theta_1 / M),
    \label{Eq:ApproximateDipoleField}
\end{equation}
where $\rho_3$ is the distance from the fiber axis, $J_1$ is the first order Bessel function of the first kind, $\theta_1 = \arcsin{\mathrm{NA}/n_1}$, with $n_1 = 1$ the refractive index before the microscope objective and $k_3 = n_1 2 \pi / \lambda $. All other contibutions vanish. 
We insert equation \eqref{Eq:ApproximateDipoleField} into equation \eqref{Eq:Overlap} and integrate numerically at different magnifications. The result can be found in Figure \ref{Fig:Overlap} b. Approximating the dipole image as a Bessel function (without any azimuthal dependence) increases the maximal coupling efficiency and shifts it to higher magnification. While qualitative behaviour remains similar, it is evident that in our configuration the approximate solution is no longer valid.
As the dipole scatterer is treated as a point source, once the light is collected by the microscope objective and imaged onto the fiber, information is distributed uniformly over the mode. For this reason the information collection efficiency will, from this point on, coincide with the photon collection efficiency.
Even though a higher NA leads to an increased information collection by the microscope, it also causes a reduced overlap of the collected light with a Gaussian single mode. Therefore it is the efficiency of the combined system that has to be considered and maximized (Figure \ref{Fig:Overlap} c). Still, computing the product of the maximal information collection efficiency $\eta_\mathrm{c}$ for each NA we notice that the overall information collection efficiency is still maximized at the highest NA.

\subsection{Objective transmission}
We measure the transmission efficiency of the microscope objective to be $\eta^*_\mathrm{d,obj}= \eta_\mathrm{d,obj} = 0.84$, assuming uniform loss, which is in good agreement with the producers specified value at this wavelength.

\subsection{Heterodyne splitting}
After mode-matching to the fiber we split 5\% by use of a variable ratio coupler of the signal to contribute to the out-of-loop heterodyne measurement (\ref{fig:setupcomplete}). We have $\eta^*_\mathrm{d,het}= \eta_\mathrm{d,het} = 0.95$.

\subsection{Homodyne balancing}
As the interferometric measurement is performed in fiber, the visibility is degraded by the imperfect splitting ratios of the variable ratio couplers. These tunable beam-splitters can be adjusted to a mismatch of about $0.1\%$, with thermal fluctuations of less than $0.5\%$. This results in an efficiency $\eta^*_\mathrm{d,hom}= \eta_\mathrm{d,hom} = 0.99$.

\subsection{Detector efficiency}
Together with the microscope transmissivity this is the second largest loss. We use a commercial balanced detector, where the current difference between the 2 diodes is amplified by a transimpedance gain. We calibrate the detector responsivity defined as $R(\nu) = \eta^*_\mathrm{d,q} e/h\nu$ with $e$ the electron charge, by measuring the dc voltage at each diode monitor port and extrapolate the efficiency of $\eta_\mathrm{d,q}^*= \eta_\mathrm{d,q} = 0.85$ for both diodes.

\subsection{Detector dark noise}
The last detection noise source is the detector dark noise. We measure the dark noise at the relevant frequencies to be $11~\mathrm{dB}$ below the shot noise level, resulting in $\eta_\mathrm{d,dn} = 0.924$.

\subsection*{Digital noise}
After detection there are further noise sources to be considered which reduce the collected information: The Red-Pitaya board has 14-bit analog to digital and digital to analog converters. This results in a limitation of the dynamic range of operation. In our settings this results in an effective information loss of $2\%$ ($\eta_\mathrm{d,rp} = 0.98$).

\subsection{Environmental information loss}
We here consider the information loss to interactions with gas molecules. This contributes the dominant environmental loss in $\eta_\mathrm{e}$. As already discussed in Section \ref{sec:theorysql}, a gas molecule colliding with the particle performs a measurement which information we cannot read. The associated efficiency is:
\begin{equation}
    \eta_\mathrm{e} = \frac{S_{F}^\mathrm{ba}}{S_{F}^\mathrm{tot}} = 0.97
\end{equation}
Values for the force noise contributions are calculated in Section \ref{sec:forcenoise}. As discussed in Section \ref{sec:forcenoise} we can define the cooperativity $C_q = S_{F}^\mathrm{ba}/S_{F}^\mathrm{th} = \Gamma_{\mathrm{ba}}/\Gamma_{\mathrm{th}}$:
\begin{equation}
    \eta_\mathrm{e} = \left(1+\frac{1}{C_q}\right)^{-1}
\end{equation}

\subsection{The total loss budget}

We finally derive a total photon detection efficiency of $\eta^* = 0.178$ while the total information detection efficiency is as high as $\eta = 0.347$. This estimation of the total information collection efficiency is in excellent agreement (less than 1\% unaccounted for) with the value of $\eta = 0.342$ directly calculated from the ratio of measurement to decoherence rates.
\begin{table}[!htb]
\centering
\setlength{\tabcolsep}{10pt} 
\renewcommand{\arraystretch}{1.5} 
\begin{tabular}{|  c  |  c  |  c  | } 
\hline
Loss source & $\eta^*$ & $\eta$ \\
\hline
\hline
Microscope collection (d) & 0.375 & 0.84 \\ 
Microcope transimissivity (d) & 0.84 & 0.84 \\
Confocal mode-matching (d) & 0.71 & 0.71 \\ 
Heterodyne split (d) & 0.95 & 0.95 \\ 
Homodyne balancing (d) & 0.99 & 0.99 \\
Detector efficiency (d) & 0.85 & 0.85 \\
Detector dark-noise (d) & - & 0.92 \\
Kalman digital noise (d) & - & 0.98 \\
Environmental information loss  (e)& - & 0.96 \\
\hline
Total  & 0.178 & 0.347 \\
\hline
\end{tabular}
\caption{\textbf{Measurement efficiency}. The total efficiency budget for photon and information loss. All loss sources are considered in both the detection and electronic line (d), and information loss to the environment (e).}
\end{table}

\section{Contributions to the total force noise}\label{sec:forcenoise}
We here estimate the expected force noise contributions given the parameters of our system. Actual values are measured in Section \ref{sec:parameteridentification}. While the backaction and thermal force noise contributions are defined and fixed by the physical system, the contribution from the feedback strongly depend on the chosen control algorithm. 

\subsection{Backaction force noise}

The backaction force noise, resulting from photons scattering off the particle was derived in Section \ref{sec:theorysql}. In order to estimate its contribution, we must consider the experimental details of the optical tweezer. The power scattered by the particle is $P_\mathrm{scatt} = I_0 \sigma$, where $I_0$ is the tweezer intensity and $\sigma =\frac{8\pi}{3}(\frac{\alpha k^2}{4\pi\epsilon_0})^2$ the scattering cross section ($\alpha$: polarizability of the particle $\epsilon_0$: vacuum permittivity). The tweezer intensity $I_0 = 2 P/\pi w$ depends on the trapping power $P$ and on the effective beam waist $w$ calculated for a tightly focused beam at the particle position, taking into account the displacement due to the scattering force contribution~\cite{Novotny}. We calculate a scattered power by the dipole of $P_\mathrm{scatt} = \SI{22.4}{\micro\watt}$. The expected single-sided backaction force noise therefore is:
\begin{equation}\label{eq:backActionNoise}
S_{F}^\mathrm{ba} = 2 \left(A^2+\frac{2}{5} \right)\frac{\hbar k P_\mathrm{scatt}}{c} = 8.4\cdot 10^{-41}\,\mathrm{N^2/Hz},   
\end{equation}
In the absence of feedback center of mass motion of the particle would thermalize to a temperature defined by competing effects of photon recoil heating and radiation damping~\cite{Novotny2017}: $T_{\mathrm{opt}}= \hbar \omega_0/(4k_{\mathrm{B}})$ ($\omega_0$: the optical laser frequency). This is equivalent to $n_\mathrm{ba} = 6.8\cdot 10^8$ quanta of occupation of the harmonic oscillator. We cannot directly observe this in the experiment as it would lead to  the particle loss due to the finite optical trap depth.

\subsection{Thermal force noise}

The thermal force noise is the noise contribution arising from interaction with the surrounding gas. At a temperature $T$ of $292~\mathrm{K}$ and pressure of $10^{-8}~\mathrm{mbar}$, we calculate:
\begin{equation}\label{eq:thermalNoise}
S_{F}^\mathrm{th} = 4 k_\mathrm{B}T\gamma_\mathrm{th} m = 3.9\cdot 10^{-42} \, \mathrm{N^2/Hz}
\end{equation}
where $k_\mathrm{B}$ is the Boltzmann constant and $\gamma_\mathrm{th}$ is the damping due to residual gas molecules (for definition see also Section \ref{sec:parameteridentification}). This force noise contributes to an occupancy of $n_\mathrm{th} = 6.0 \cdot 10^7$.

\subsection{Feedback force noise}
\begin{figure*}[!h]
\includegraphics[scale=1]{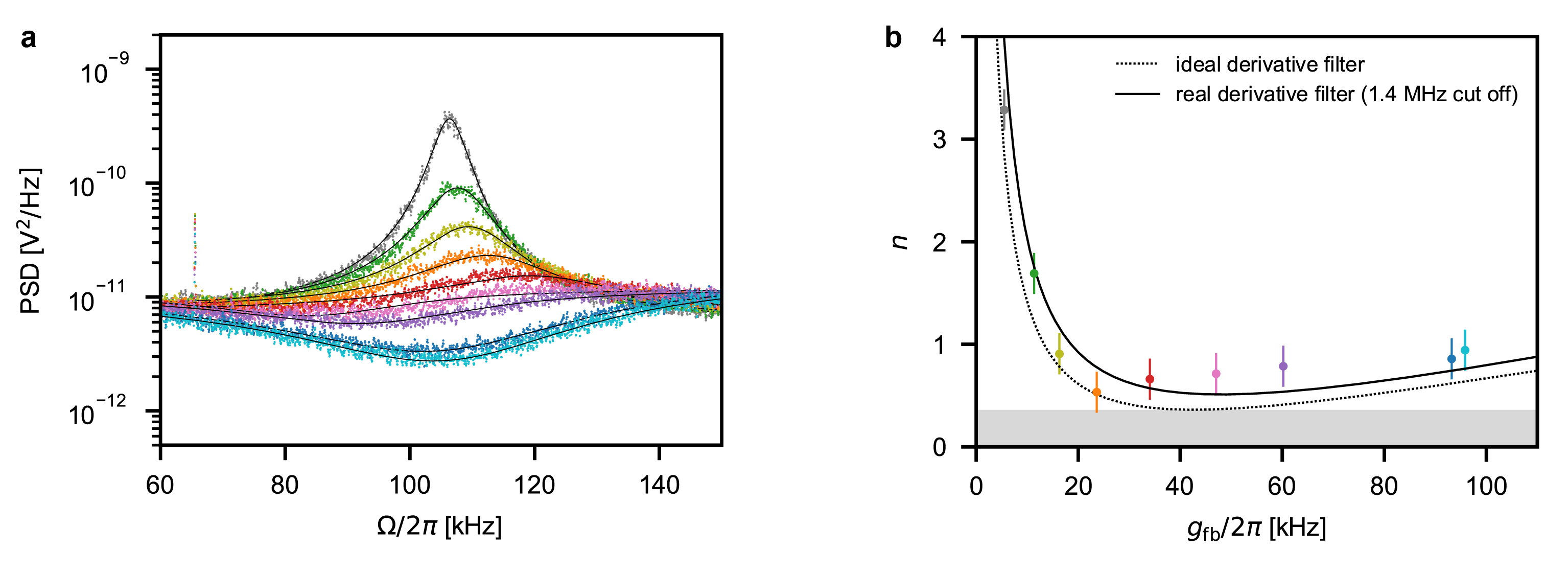}
\caption{\textbf{Derivative feedback performance}. \textbf{a}, Noise squashing in the closed-loop measurement PSD resulting from leakage of measurement noise inthe colsed-loop system. \textbf{b} Occupation measured in heterodyne detection as a function of the feedback gain of a derivative feedback. The different colours cossepond to the PSDs in \textbf{a}. A qualitative approximation of this behaviour can be obtained by tuning the Kalman gains (defined in Section \ref{sec:kalman}) of our controller to a value that is a factor $10^5$ larger than the optimal one, and reducing the controller transfer function to be white over a large frequency band. In this setting the filter ignores the model and bases its feedback solely on the measurement. It is important to notice that the ideal differentiation filter would reach occupations as low as those determined by the measurement uncertainty as defined in \eqref{eq:nmin}. However this is not a practical solution, as it would require an infinite bandwidth controller.}
\label{fig:squashing}
\end{figure*}

Measurement-based feedback control relies on a typically noisy measurement to control the dynamics of the system. The measurement noise is therefore fed back to the controller whose output drives the system, adding a new contribution to the force noise term, and setting a lower bound to the accuracy of the control.  
The force noise arising from feedback imprecision noise is:
\begin{equation}
S_{F}^\mathrm{fb}(\Omega) = \lvert h_\mathrm{fb}(\Omega)\rvert^2 S_{z}^\mathrm{imp},
\label{eq:fbff}
\end{equation}
where $ h_\mathrm{fb}(\Omega)$ is the controller transfer function in the feedback path. Closing the feedback loop the susceptibility of the system becomes:
\begin{equation}
    \chi_\mathrm{eff}(\Omega) = \frac{\chi_\mathrm{m}(\Omega)}{1+\chi_\mathrm{m}(\Omega)h_\mathrm{fb}(\Omega)},
\label{eq:chieff}
\end{equation}
which allows us to write the closed-loop spectral density of the position ($z$) and measurement outcome ($\zeta$):
\begin{subequations}
\begin{align}
    S_{z}(\Omega) &=  \lvert \chi_\mathrm{eff}(\Omega)\rvert^2\left(S_{F}^\mathrm{tot} + \lvert h_\mathrm{fb}(\Omega)\rvert^2 S_{z}^\mathrm{imp}\right)\\
   S_{\zeta}(\Omega) &=  \lvert \chi_\mathrm{eff}(\Omega)\rvert^2\left(S_{F}^\mathrm{tot} + \lvert \chi_m(\Omega)\rvert^{-2} S_{z}^\mathrm{imp}\right)
\end{align}
\label{eq:szfb}
\end{subequations}
From  (\ref{eq:fbff}),(\ref{eq:chieff}) and (\ref{eq:szfb}a), we see how the controller transfer function $h_\mathrm{fb}(\Omega)$ influences the closed-loop power spectral densities (PSDs). The controller should minimize the PSD by respecting the constraints of the control input and render the closed-loop stable. For linear Gaussian systems such as the one considered in this paper, the linear-quadratic Gaussian (LQG) controller fulfills these demands in an optimal way, as will be discussed in detail in the next sections. We here discuss the simple example of a \textit{differentiation filter}, as it is the most common form of feedback cooling applied in most optomechanical experiments.
The feedback transfer function for the differentiation filter is:
\begin{equation}
 h_\mathrm{fb}^{d}(\Omega) = \ii m \Omega \gamma g_\mathrm{fb} 
\label{eq:tfunc}
\end{equation}
where $\gamma$ is the natural damping of the system associated to the bath (of temperature $T$) it is coupled to, and $g_\mathrm{fb}$ the feedback gain. Evaluating the total energy as a function of the gain $g_\mathrm{fb}$ makes it evident that at some point the imprecision noise will start to be fed back into the system, heating the motion of the particle:
\begin{equation}
    \left\langle z^2 \right\rangle = \frac{1}{2\pi} \int_0^\infty S_z(\Omega) d\Omega = \frac{1}{1+g_\mathrm{fb}} \frac{k_\mathrm{B} T}{m\Omega_z^2} + \frac{g_\mathrm{fb}^2}{1+g_\mathrm{fb}}\frac{\gamma}{2} \frac{S_{z}^\mathrm{imp}}{2}.
\label{eq:derivative}
\end{equation}
This effect appears in the measured spectral density $S_{\zeta}(\Omega)$ in the form of noise squashing, as the particle motion is driven to minimize the total noise in the measurement outcome~\cite{Poggio2007,Rossi2018, Wilson2015}.
It is important to notice that, in practical applications, the controller transfer function defined in \eqref{eq:tfunc} is not realistic as exact differentiation would require infinite bandwidth and knowledge of the future, producing unbounded control signals. When limiting the bandwidth of the differentiation filter, the qualitative behaviour of \eqref{eq:derivative} is preserved, albeit with a reduced performance (see Figure \ref{fig:squashing}b).

\subsection{Coupling of the transverse degrees of freedom}

The finite temperature of the transverse modes may in principle affect the cooling performance in the $z$-direction in two ways:
\begin{itemize}
    \item 	Coupling of the transverse degrees of freedom through the measurement. In this case, displacements along the transverse directions are transduced into the backscattered signal. This effect would reduce the information efficiency of the $z$-measurement, just as any other noise source, in turn reducing the cooling performance. For a specific measurement geometry, the noise power contributed by each mode $i=(x,y,z)$ can be written as $P_i\propto \Gamma_{i}^{\mathrm{meas}} \left(\langle2 n_{i} \rangle+1\right)$, where $\Gamma_{i}^{\mathrm{meas}}$ are the measurement rates for each mode, and $\langle n_{i} \rangle$ the average occupation of each mode. Concretely, for our setup, we find  $\Gamma_{x,y}^{\mathrm{meas}}/\Gamma_{z}^{\mathrm{meas}}\sim ~10^{-5}$, which means that the residual coupling of the transverse modes is about 5 orders of magnitude weaker than for the z-mode. Using parametric feedback via an independent forward detection scheme (Figure S1), the transverse modes are cooled to occupations of  $\langle n_{x,y} \rangle \sim 10^3$. This yields a relative noise power contribution of the two transverse modes, when $\langle n_{z} \rangle \sim 1$, of about  $P_{x,y}/P_z \sim 10^{-2}$ , which is a negligible contribution. In addition, since the feedback signal for cooling is confined to the spectral region around $\Omega_z$, the spectral separation between transverse motion and z-motion further suppresses the unwanted cross-coupling effect.

    \item Coupling between the modes through the nonlinearity of the potential. The optical tweezer presents a duffing nonlinearity of the order $\xi_i=-2/w_i^2$ with $i=(x,y,z)$ and $w_i$ the beam characteristic length scale (i.e. the waist for the transverse directions (x,y) and the Rayleigh length  for the z-direction). As a consequence, the force along each direction of motion becomes coupled to the position in the other directions~\cite{Gieseler2013}:
    \begin{equation}
    F_i= -k_i x_i \left(1+ \sum_{j=x,y,z}\xi_j x_j^2 \right).
    \end{equation}
    For small displacements, $\lvert x \rvert \ll \lvert \xi_i\rvert^{-1/2}$, this coupling becomes negligible and the modes decouple. Specifically, in this experiment, we have $\vert \xi_i\rvert^{-1/2} \geq 4 \times 10^{-7} \,\mathrm{m}$ and the root mean square displacement along each direction is given by $x_i^{\mathrm{rms}} = x_i^{\mathrm{zpf}}\sqrt{2n_i+1}$. While the $z$-motion is cooled to   $\langle n_{z} \rangle \sim 0.5$, the motion along the other modes is parametrically cooled to $\langle n_{x,y} \rangle \sim 10^3$.  This is enough to have $x_{x,y}^{rms} \sim 10^{-10}\,\mathrm{m}$. It is evident that even with limited cooling on the transverse modes the expected energy contribution to the z-mode due to nonlinear coupling is negligible. 

\end{itemize}

\section{Quantum equations of motion}
\label{sec:qlangevin}
In this section we derive the quantum Langevin equations for the nanosphere, describing its motion in the harmonic trap formed by the tweezer field, together with the corresponding input--output relations. These equations form the basis for the state-space model used for the Kalman filter.

\subsection{Hamiltonian}
The effective Hamiltonian for the center-of-mass motion of the nanosphere in the tweezer field and the coupling to the electromagnetic field can be derived following~\cite{gonzalez-ballestero_theory_2019} (which treats a more general system),
assuming a linear, isotropic dielectric medium and the validity of the
long-wavelength assumption. (That is, the typical extension of the mechanical
state is much smaller than the tweezer wavelength $\lambda_0$).

We describe the center-of-mass motion of the nanosphere in direction
$j\in\{x,y,z\}$ by annihilation and creation operators $\bop_j$ and $\bdag_j$
with commutation relations $[\bop_i,\bdag_j]=\delta_{ij}$. The light field is
expanded into a continuum of plane-wave modes labeled by the wavevector $k\in \mathds{R}^3$ and an index $\lambda$ that determines the mode's polarisation.
The corresponding annihilation and creation operators are denoted by
$\aop_{\lambda}(k)$ and $\adag_{\lambda}(k)$, respectively. Their commutation relations are given by
$[\aop_{\lambda}(k),\adag_{\lambda'}(k')]=\delta_{\lambda\lambda'}\delta(k-k')$,
where $\delta(\cdot)$ is the Dirac $\delta$-function and
$\delta_{\lambda\lambda'}$ is the Kronecker $\delta$. For the system discussed
here we find the Hamiltonian
\begin{equation}
  \label{eq:qle-qle-1}
  H=\hbar\sum_{j=x,y,z}\Omega_j\bdag_j\bop_j+\hbar\sum_{\lambda}\int\ddd{k}\Delta_k\adag_{\lambda}(k)\aop_{\lambda}(k)+\hbar\sum_{j=x,y,z}\sum_{\lambda}\int \ddd{k}\left[ g_{j\lambda}(k)\adag_{\lambda}(k)(\bop_j+\bdag_j) + \mathrm{H.c.}\right],
\end{equation}
where $\Omega_j$ is the mechanical frequency in direction $j$,
$\Delta_k=\omega_k-\omega_0$, and $\omega_k=\lVert{k}\rVert c$. The coupling constants $g_{j\lambda}(k)$ are
given by
\begin{align}
  \label{eq:qle-2}
  g_{j\lambda}(k)&=\ii \frac{G_0^{\lambda}(k)}{2}(k_j-k_0\delta_{jz})r_{0j},\\
  G_0^{\lambda}(k)&=\alpha E_0 \sqrt{\frac{\omega_k}{2 \hbar \varepsilon_0 (2\pi)^3}}\ee_x\cdot \ee_{\lambda}(k),
\end{align}
where $r_{0j}$ is the mechanical ground-state extension in direction $j$,
$\alpha$ is the nanosphere's polarisability, and $E_0$ is the electric field
strength of the tweezer. Symbols $\ee_x$ and $\ee_{\lambda}(k)$ denote unit
vectors in $x$-direction and the direction of (linear) polarization for the $(k,\lambda)$
field mode respectively, and $\ee_x\cdot \ee_{\lambda}$ denotes their scalar
product in $\mathds{R}^3$.

\subsection{Quantum Langevin Equations}
Starting from the Hamiltonian above we now derive the quantum-optical Langevin equations for the mechanical system following the procedure introduced in~\cite{gardiner_input_1985}. Here we neglect relativistic effects \cite{Novotny2017} and, for now, also mechanical damping effects due to residual gas which will be added later. We first find the Heisenberg equations for $\bop_j$ and $\aop_{\lambda}(k)$, yielding
\begin{subequations}
  \begin{align}
    \label{eq:qle-3a}
    \dot{\aop}_{\lambda}(k,t)&=
                               -\ii\Delta_k\aop_{\lambda}(k,t)-\ii \sum_jg_{j\lambda}(k)(\bop_j+\bdag_j),\\
    \label{eq:qle-3b}
    \dot{b}_j(t)&=
                  -\ii\Omega_j\bop_j(t)-\ii\sum_\lambda\int \ddd{k}[g_{j\lambda}(k)\adag_{\lambda}(k,t)+\mathrm{H.c.}].
  \end{align}
\end{subequations}
We formally solve \eqref{eq:qle-3a}, which gives
\begin{equation}
  \label{eq:qle-3}
  \aop_{\lambda}(k,t)=\ee^{-\ii \Delta_k t}\aop_{\lambda}(k,0)-\ii \sum_{j=x,y,z}g_{j\lambda}(k)\int_0^t\dd{s}\ee^{-\ii\Delta_k(t-s)}[\bop_j(s)+\bdag_j(s)],
\end{equation}
and plug it into \eqref{eq:qle-3b}. We find
\begin{multline}
  \label{eq:qle-4}
  \dot{\bop}_j(t)=-\ii\Omega_j\bop_j(t)-\ii\sum_{\lambda}\int\ddd{k}[g_{j\lambda}(k)\ee^{\ii\Delta_k
    t}\adag_{\lambda}(k,0)+\mathrm{H.c.}]\\+\sum_{l=x,y,z}\int_0^t\dd{s}[\bop_j(s)+\bdag_j(s)]\int\ddd{k}\sum_{\lambda}[g_{j\lambda}(k)g_{l\lambda}^{*}(k)\ee^{\ii
    \Delta_k(t-s)}-\mathrm{H.c.}].
\end{multline}
We now make the (typical) assumptions \cite{reynaud_quantum_1997} that (\hypertarget{approx1}{i}) the interaction with
the field is restricted to a frequency interval
$[\omega_0-\theta,\omega_0+\theta]$, where $\theta$ is a cutoff frequency that
fulfills $\omega_0\gg \theta \gg \Omega_j$, and (\hypertarget{approx2}{ii}) the coupling constants
$g_{j\lambda}$ are approximately constant across this interval. These
assumptions will allow us to employ a Markov approximation (taking the limit $\theta\rightarrow\infty$), making the resulting
equation local in time, and considerably simplify equation \eqref{eq:qle-4}.

We first take a look at the second term in (\ref{eq:qle-4}), which describes the
interaction of the mechanical system with (unnormalized) light modes
$\int\ddd{k}g_{j\lambda}^{*}(k)\ee^{-\ii \Delta_k t}\aop_{\lambda}(k,0)$, where
$t$ should be interpreted as the time at which the incoming light-field
interacts with the nanosphere. For our purposes it is convenient to decompose
this mode into two orthogonal modes, one of which is monitored in the
experiment. The corresponding mode function, denoted by $h$, is determined by
the measurement setup. 
We write
\begin{equation*}
  \int\ddd{k}g_{l\lambda}^{*}(k)\ee^{-\ii \Delta_k t}\aop_{\lambda}(k,0)=
  \sqrt{2\pi K_{ll}^{\lambda}}\left\{ \sqrt{\eta_{l\lambda}}c_{\lambda}(t)+\sqrt{1-\eta_{l\lambda}}c_{l\lambda}^{\perp}(t) \right\},
\end{equation*}
where we defined the light modes
\begin{subequations}
  \begin{align}
    \label{eq:qle-6}
    c_{\lambda}(t)&=(2\pi I)^{-\frac{1}{2}}\int\ddd{k}h^{*}(k)\ee^{-\ii \Delta_k  t}\aop_{\lambda}(k,0),\\
    c_{l\lambda}^{\perp}(t)&=[2\pi K_{ll}^{\lambda}(1-\eta_{l\lambda})]^{-\frac{1}{2}}\int\ddd{k}[g_{l\lambda}^{*}(k)-(J^*_{l\lambda}/I) h^{*}(k)]\ee^{-\ii \Delta_k  t}\aop_{\lambda}(k,0),
  \end{align}
\end{subequations}
and the constants
\begin{align}
  \label{eq:qle-7}
  I&=\int\dd{\Omega_{k}}\frac{\omega_0^2}{c^3}\left\lvert h\!\left(\frac{\omega_0}{c}\ee_k\right)\right\rvert^{2},\\
  J_{l\lambda}&=\int\dd{\Omega_{k}}\frac{\omega_0^2}{c^3}h^{*}\left(\frac{\omega_0}{c}\ee_k\right)g_{l\lambda}\left(\frac{\omega_0}{c}\ee_k\right),\\
  K_{jl}^{\lambda}&=\int\dd{\Omega_{k}}\frac{\omega_0^2}{c^3}g_{j\lambda}\left(\frac{\omega_0}{c}\ee_k\right)g_{l\lambda}^{*}\left(\frac{\omega_0}{c}\ee_k\right).
\end{align}
Here $\dd{\Omega_{k}}$ denotes the integration with respect to the angular
degrees of freedom of $k$ and $\ee_k$ is a unit vector in the direction of $k$.
The parameter $\eta_{l\lambda}=|J_{l\lambda}|^2/ K_{ll}^{\lambda}I \in [0,1]$
determines the overlap between the measured mode function $h$ and the scattering
profile $g_{l\lambda}$ at the tweezer frequency $\omega_0$ and takes the role of
a measurement efficiency. Note that for $h=g_{l\lambda}$ we have
$\eta_{l\lambda}=1$. The parameter $K_{ll}^{\lambda}$ on the other hand
effectively describes the coupling strength between the nanosphere's motion in
direction $l$ and the mode light mode defined by $g_{l\lambda}$. Plugging the
expressions for $g_{l\lambda}$ into the definition of $K_{jl}^{\lambda}$ one can
show that $K_{jl}^{\lambda}=K_{ll}^{\lambda}\delta_{jl}$.

Assuming that $h$ is (similarly to $g$) restricted to a
frequency interval around $\omega_0$ and roughly flat and taking the Markovian
limit ($\theta \rightarrow \infty$) we can show that $c_{\lambda}$,
$c_{\lambda}^{\perp}$ describe zero-mean white-noise fields that obey
\begin{subequations}
\label{eq:qle-comm-inputs}
  \begin{align}
  \label{eq:qle-8}
  [c_{\lambda}(t),c^{\dagger}_{\lambda'}(s)]&=[c_{l\lambda}^{\perp}(t),(c^{\perp}_{l\lambda'})^{\dagger}(s)]=\delta_{\lambda\lambda'}\delta(t-s),\\
  [c_{\lambda}(t),(c^{\perp}_{l\lambda'})^{\dagger}(s)]&=0,
\end{align}
\end{subequations}
and, assuming the electromagnetic field is initially in the vacuum state, the correlation functions
\begin{subequations}
  \label{eq:qle-corr-funcs}
  \begin{align}
  \langle c_{\lambda}(t)c^{\dagger}_{\lambda'}(s)\rangle &=\langle c_{l\lambda}^{\perp}(t)(c^{\perp}_{l\lambda'})^{\dagger}(s)\rangle=\delta_{\lambda\lambda'}\delta(t-s),\\
  \langle c_{\lambda}(t)(c^{\perp}_{l\lambda'})^{\dagger}(s)\rangle &=0,
\end{align}
\end{subequations}
where $\langle \cdot \rangle$ refers to the expectation value with respect to system plus environment. In deriving relations \eqref{eq:qle-comm-inputs} and \eqref{eq:qle-corr-funcs} we find integrals of the following form,
which can be approximated using the assumptions (\hyperlink{approx1}{i}) and (\hyperlink{approx2}{ii}) from above:
\begin{align}
\label{eq:qle-maap}
  \int\ddd{k}g_{j\lambda}(k)g_{l\lambda}^{*}(k)\ee^{\ii \Delta_k(t-s)}
    &\overset{(i),(ii)}{\approx}\int_{\omega_0-\theta}^{\omega_0+\theta}\dd{\omega}\ee^{\ii (\omega-\omega_0)(t-s)}\int\dd{\Omega_{k}}\frac{\omega_0^2}{c^3}g_{j\lambda}\left(\frac{\omega_0}{c}\ee_k\right)g_{l\lambda}^{*}\left(\frac{\omega_0}{c}\ee_k\right)\\
    &\underset{\theta\rightarrow \infty}{\longrightarrow} 2\pi K_{jl}^{\lambda}\delta(t-s)\nonumber
\end{align}
Plugging this back
into equation \eqref{eq:qle-4} we see that, under this approximation, the second
line vanishes identically as $K_{lj}^{\lambda}\in \mathds{R}$.
Using this, the quantum Langevin equations for the motion of the nanosphere (in
a Markov approximation) take the form
\begin{equation}
  \label{eq:qle-9}
  \dot{\bop}_l(t)=-\ii\Omega_l\bop_l(t)-\ii \sum_{\lambda}\sqrt{2\pi K_{ll}^{\lambda}}\left\{ \sqrt{\eta_{l\lambda}}[c_{\lambda}(t)+c_{\lambda}^{\dagger}(t)]+\sqrt{1-\eta_{l\lambda}}[c_{\lambda}^{\perp}(t)+(c_{\lambda}^{\perp}(t))^{\dagger}] \right\}.
\end{equation}
Alternatively we can rewrite (\ref{eq:qle-9}) in terms of position
$r_j=(\bop_j+\bdag_j)/\sqrt{2}$ and momentum $p_j=(\bop_j-\bdag_j)/\sqrt{2}\ii$
\begin{subequations}
  \label{eq:qle-10}
  \begin{align}
    \dot{r}_l(t)&=\Omega_lp_l(t),\\ %
    \dot{p}_l(t)&=-\Omega_lr_l(t)-\sum_{\lambda}\sqrt{4\pi
               K_{ll}^{\lambda}}\left\{
               \sqrt{\eta_{l\lambda}}x_{\lambda}(t)+\sqrt{1-\eta_{l\lambda}}x_{l\lambda}^{\perp}(t)
               \right\},
  \end{align}
\end{subequations}
where we introduced the amplitude quadratures
$x_{\lambda}=c_{\lambda}+c_{\lambda}^{\dagger}$.

Up to now we have neglected two important points in our treatment: the nanosphere's interaction with residual gas,
which constitutes an additional thermal environment, and the feedback force. The former we model as \textit{Brownian motion damping}~\cite{gardiner_quantum_2004},
but treat it in a Markov
approximation. We thus introduce an additional Gaussian noise operator $f_l$ that
obeys
\begin{subequations}
  \label{eq:qle-11}
  \begin{align}
    \mean{f_l(t)}&=0,\\
    \mean{f_{l}(t)f_{l}(t')+f_l(t')f_l(t)}&=(2\bar{n}_l+1)\delta(t-t'),
  \end{align}
\end{subequations}
where $\bar{n}_l=\hbar \Omega_l/k_{\mathrm{B}}T$. The corresponding damping rate we denote by $\gamma$. The additional energy contribution by the feedback we write as $H_{\mathrm{fb}}=-qE_{\mathrm{fb}}(t)r_{0z}r_z=-\hbar u(t)r_z$, where $q$ is the charge of the particle and $E_{\mathrm{fb}}(t)$ is the time-dependent electric field that is used to apply the feedback signal (also see Section \ref{sec:parameteridentification}). Putting this all together the modified Langevin equations take the form
\begin{subequations}
  \label{eq:qle-12}
  \begin{align}
    \label{eq:qle-13}
    \dot{r}_l(t)&=\Omega_lp_l(t),\\ %
    \label{eq:qle-13b}
    \dot{p}_l(t)&=-\Omega_lr_l(t)-\gamma p_l(t)+u(t)+\sqrt{2\gamma}f_l(t)-\sum_{\lambda}\sqrt{4\pi
               K_{ll}^{\lambda}}\left\{
               \sqrt{\eta_{l\lambda}}x_{\lambda}(t)+\sqrt{1-\eta_{l\lambda}}x_{l\lambda}^{\perp}(t)
               \right\}.
  \end{align}
\end{subequations}
A relativistic treatment of the optomechanical interaction would as well show a radiation-damping contribution to the particle dynamics~\cite{Novotny2017}.
Together with the radiation-pressure shot noise (described by the last term in \eqref{eq:qle-13b}) this defines, similarly to the thermal environment, a fluctuation--dissipation balance and a thermalization temperature associated with the optical bath. In our experiment both damping mechanisms (residual gas and radiation damping) are negligible in the presence of feedback. The experimental decoherence rates for the thermal and optical interactions are characterized in Section~\ref{sec:parameteridentification}.

\subsection{Input--Output relations}
To compute the scattered field after the interaction with the nanosphere (that
is what we measure) we go
back to equation (\ref{eq:qle-3}) which, in a first step, we multiply by $h^{*}(k)$ and
integrate over $k$, leading to
\begin{align}
  \label{eq:qle-14}
  \int \ddd{k}h^{*}(k)\aop_{\lambda}(k,t)&=\sqrt{2\pi I}c_{\lambda}(t)-\ii \sum_{l=x,y,z}\int \ddd{k}h^{*}(k)g_{l\lambda}(k)\int_0^t\dd{s}\ee^{-\ii\Delta_k(t-s)}[\bop_l(s)+\bdag_l(s)]\nonumber\\
                                         &\approx\sqrt{2\pi I}c_{\lambda}(t)-2\ii\pi \sum_{l=x,y,z} J_{l\lambda}\int_0^t\dd{s}\delta(t-s)[\bop_l(s)+\bdag_l(s)]\nonumber\\
                                         &=\sqrt{2\pi I}c_{\lambda}(t)-\ii \pi  \sum_{l=x,y,z}J_{l\lambda}[\bop_l(t)+\bdag_l(t)]
\end{align}
Note again that $c_{\lambda}(t)$ refers to the light field \textit{before} the
interaction. To connect this to its state \textit{after} the interaction, we
again formally integrate (\ref{eq:qle-3a}), this time specifying $a(k,T)$ at some
(distant) final time $T>t$:
\begin{equation}
  \label{eq:qle-15}
  \aop_{\lambda}(k,t)=\ee^{\ii \Delta_k (T-t)}\aop_{\lambda}(k,T)+\ii \sum_{j=x,y,z}g_{j\lambda}(k)\int_t^T\dd{s}\ee^{-\ii\Delta_k(t-s)}[\bop_j(s)+\bdag_j(s)].
\end{equation}
Applying the same procedure as before we find
\begin{equation}
  \label{eq:qle-16}
  \int \ddd{k}h^{*}(k)\aop_{\lambda}(k,t)= \sqrt{2\pi I}c_{\lambda}^{\mathrm{out}}(t)+\ii \pi  \sum_{l=x,y,z}J_{l\lambda}[\bop_l(t)+\bdag_l(t)],
\end{equation}
where we interpret $c_{\lambda}^{\mathrm{out}}(t)=\int\ddd{k}h^{*}(k)\ee^{\ii \Delta_k
  (T-t)}\aop_{\lambda}(k,T)/\sqrt{2\pi I}$ as the field (at the time $T$) after the
interaction. We can combine equations (\ref{eq:qle-16}) and (\ref{eq:qle-14}) to obtain
the usual input--output relation (with $\varphi_{l\lambda}=\arg J_{l\lambda}$)
\begin{equation}
  \label{eq:qle-17}
  c_{\lambda}^{\mathrm{out}}(t)=c_{\lambda}(t)-\ii  \sum_{l=x,y,z}\sqrt{2\pi \eta_{l\lambda}K_{ll}^{\lambda}}\ee^{\ii \varphi_{l\lambda}}[\bop_l(t)+\bdag_l(t)].
\end{equation}
Note that the choice of $h$ and thus the value of $\eta_{l\lambda}$ determines
which direction of the nanosphere's motion can be monitored by measuring the
scattered light.
In the experiment we use homodyne detection to monitor (amplitude and phase)
quadratures ($x_{j\lambda}^{\mathrm{out}}$ and $y_{j\lambda}^{\mathrm{out}}$) of
the scattered field. The corresponding input--output relations are given by
\begin{subequations}
  \label{eq:qle-18}
  \begin{align}
    \label{eq:qle-19}
    x_{\lambda}^{\mathrm{out}}(t)&=[c_{\lambda}^{\mathrm{out}}(t)+(c_{\lambda}^{\mathrm{out}}(t))^{\dagger}]=x_{\lambda}(t)+\sum_{l=x,y,z}\sin \varphi_{l\lambda}\sqrt{16\pi \eta_{l\lambda}K_{ll}^{\lambda}}r_{l}(t),\\
    \label{eq:qle-19b}
    y_{\lambda}^{\mathrm{out}}(t)&=-\ii[c_{\lambda}^{\mathrm{out}}(t)-(c_{\lambda}^{\mathrm{out}}(t))^{\dagger}]=y_{\lambda}(t)- \sum_{l=x,y,z}\cos\varphi_{l\lambda}\sqrt{16\pi \eta_{l\lambda}K_{ll}^{\lambda}}r_{l}(t).
  \end{align}
\end{subequations}
As in our experiment $\varphi_{z\lambda}\approx 0$ the amplitude quadrature $x_{\lambda}^{\mathrm{out}}$ only
carries noise, while the phase quadrature $y_{\lambda}^{\mathrm{out}}$ contains
information about the nanosphere's position. We thus only
monitor the phase quadrature. Also, (\ref{eq:qle-19b}) shows that, depending on the value of $\eta_{l\lambda}$ and thus on the definition of the measured mode $h$, $y_\lambda^{\mathrm{out}}$ contains contributions from the particle displacement along all directions. In the experiment $h$ is such that the contributions from the $x$ and $y$ directions are heavily suppressed (i.e., $\eta_{x\lambda},\eta_{y,\lambda}\ll\eta_{z,\lambda}$). Additional imperfections in the experimental setup will determine the effective measurement efficiency, which will result in effective values for $\eta_{l\lambda}$ (see Section \ref{sec:losses}).

\subsection{Quantum Langevin equations in vector form}
\label{sec:qlangevin-vector}
In analogy to the \emph{state-space models} commonly used in classical control theory, we can rewrite the quantum Langevin equations \eqref{eq:qle-12} and the input--output relations \eqref{eq:qle-18} in vector form. These definitions will enable us to compactly write the Kalman filter equations in the next section.

We start by defining $\mathbf{z}(t)=[r_z(t) \enspace p_z(t)]^{\mathrm{T}}$.
Here and in the following sections, we assume that we measure the phase quadrature $y_{\lambda_0}^{\mathrm{out}}(t)$ for a single polarisation $\lambda_0$. We can then write
\begin{subequations}
  \label{eq:contStateSpace}
  \begin{align}
    \label{eq:contStateSpace-z}
    \dot{\mathbf{z}}(t)&=\mathbf{A} \mathbf{z}(t) + \mathbf{b}u(t) + \mathbf{w}(t),\\
    \label{eq:contStateSpace-y}
    y_{\lambda_0}^{\mathrm{out}}(t) &= \mathbf{c}^{\mathrm{T}} \mathbf{z}(t) + y_{\lambda_0}(t),
  \end{align}
\end{subequations}
with
\begin{align}
  \mathbf{A}&=\begin{bmatrix}
    0&\Omega_z\\
    -\Omega_z&-\gamma
  \end{bmatrix},
  &\mathbf{b}=\begin{bmatrix} 0&1\end{bmatrix}^T,&
  &\mathbf{c}^{\mathrm{T}}&= \sqrt{16\pi \eta_{z\lambda_{0}}K_{zz}^{\lambda_{0}}}
  \begin{bmatrix} 1&0\end{bmatrix}.
\end{align}
and $\mathbf{w}(t)=\mathbf{g}w(t)=\left[0 \enspace 1\right]^T w(t)$,
\begin{equation}
    \label{eq:qle-20}
    w(t)=\left\{ \sqrt{2\gamma}f_z(t)-\sum_{\lambda}\sqrt{4\pi K_{zz}^{\lambda}}\left[
                   \sqrt{\eta_{z\lambda}}x_{\lambda}(t)+\sqrt{1-\eta_{z\lambda}}x_{z\lambda}^{\perp}(t)
                   \right] \right\}.
  \end{equation}
As the light field is assumed to be in the vacuum state both $\mathbf{w}$ and
$y_{\lambda_0}$ are zero-mean Gaussian processes. Their symmetrized (cross-)correlation
matrices are
\begin{subequations}
\label{eq:qle-corr-matrices}
  \begin{align}
    \label{eq:qle-21}
    \mean{y_{\lambda_0}(t)y_{\lambda_0}(t')}&=\delta(t-t'),\\
    \mathrm{Re}\mean{\mathbf{w}(t)y_{\lambda_0}(t')}&=\mathbf{M}\delta(t-t')=0,\\
    \mathrm{Re}\mean{\mathbf{w}(t)\mathbf{w}^{T}(t')}&=\mathbf{N}\delta(t-t')=\mathrm{diag}\Bigl(0,\gamma(2\bar{n}_z+1)+4\pi\sum_{\lambda}K_{zz}^{\lambda}\Bigr)\delta(t-t'),
  \end{align}
\end{subequations}
which follows from \eqref{eq:qle-corr-funcs} and \eqref{eq:qle-11}.

\subsection{Connection to the stochastic master equation and Kalman filtering}
\label{sec:qlangevin-kalman}
Equations (\ref{eq:qle-12}) and (\ref{eq:qle-18}) define a quantum stochastic
model of the experimental setup. This model also allows us to construct the
dynamical equations for the so-called \textit{conditional quantum state}
$\hat{\rho}$, which describes the nanosphere's motional state in $z$-direction conditioned on the classical
output of the measurement of $y_{\lambda_0}^{\mathrm{out}}$. The time evolution of $\hat{\rho}$ is (approximately) given by the Ito stochastic master equation (see, e.g., \cite{Wieseman2010} for an introduction to the formalism). Assuming $\varphi_{z\lambda_0}=0$:
 
\begin{subequations}
\begin{align}
  \label{eq:qle-sme}
    d\hat{\rho}(t)&=-\ii[\Omega_z b_z^\dagger b_z-u(t)r_z,\hat{\rho}(t)]\dd{t}+\gamma (\bar{n}+1)\mathcal{D}[b_z]\hat{\rho}(t)\dd{t}+\gamma \bar{n}\mathcal{D}[b_z^{\dagger}]\hat{\rho}(t)\dd{t}\notag\\&\qquad\qquad\qquad\qquad\qquad\qquad\qquad\qquad
    +\sum_{\lambda}\mathcal{D}[s_{z\lambda}]\hat{\rho}(t)\dd{t}+\sqrt{\eta_{z\lambda_0}}\mathcal{H}[s_{z\lambda_0}]\hat{\rho}(t)\dd{W(t)},\\
 \mathcal{D}[s]\hat{\rho}&=s\hat{\rho} s^{\dagger}-\frac{1}{2}(s^{\dagger}s\hat{\rho}+\hat{\rho} s^{\dagger}s),\\
 \mathcal{H}[s]\hat{\rho}&=[s-\mathrm{tr}(s\hat{\rho})]\hat{\rho}+\hat{\rho}[s-\mathrm{tr}(s \hat{\rho})]^{\dagger},
\end{align}
\end{subequations}
where $s_{z\lambda}=-i\sqrt{2\pi K_{zz}^\lambda}(b_z+b_z^\dagger)$. The second and third term in (\ref{eq:qle-sme}) describe damping and decoherence effects due to the residual gas, while the fourth term describes diffusion due to the coupling to the electromagnetic field. The last term effects conditioning on the homodyne measurement, where $W$ is a classical Wiener process corresponding to the innovation process denoted as $\epsilon$ in the main text. We can (formally) write for the Wiener increments $\dd{W}(t)=\epsilon(t)\dd{t}$.

In deriving this equation, we assumed that the measured mode $h$ couples only weakly to the particle motion in $x$- and $y$-direction and thus neglected measurement terms scaling with $\sqrt{\eta_{x\lambda_0}}$ and $\sqrt{\eta_{y\lambda_0}}$ (which show up as sharp resonances in the measured spectrum, see Figure~1b in main text). Also note that this formulation of mechanical damping due to residual gas does not strictly correspond to Brownian motion damping as used above. The two formulations are connected by a rotating-wave approximation (see~\cite{gardiner_quantum_2004}), which is a good approximation for oscillators with a high quality factor.

For Gaussian systems, such as ours, it was
shown \cite{Belavkin1980,edwards_optimal_2005} that the evolution of the conditional quantum state $\hat{\rho}$ can
be mapped to the well-known \textit{Kalman--Bucy filter} from classical estimation theory. In this case, $\hat{\rho}$ is completely determined by the first and second moment of $\mathbf{z}=[r_z, p_z]$ (an operator in the Schr\"odinger picture), which we denote as
\begin{subequations}
\label{eq:qle-22-both}
  \begin{align}
    \label{eq:qle-22}
    \mathbf{\hat{z}}(t)&=\mathrm{tr}[\hat{\rho}(t)\mathbf{z}],\\
    \mathbf{\hat{\Sigma}}(t)&=\mathrm{Re} \left\{ \mathrm{tr}[\hat{\rho}(t)\mathbf{z}\mathbf{z}^{T}] \right\}- \mathbf{\hat{z}}(t)\mathbf{\hat{z}}^{T}(t).
  \end{align}
\end{subequations}
Using the definitions from Section \ref{sec:qlangevin-vector} the dynamical equations determining the evolution of $\mathbf{\hat{z}}(t)$ and
$\mathbf{\hat{\Sigma}}(t)$ can be written as the classical Kalman--Bucy filter~\cite{Belavkin1980,Belavkin1998,Doherty1999,edwards_optimal_2005}
\begin{subequations}\label{eq:kalmanBucy}
  \begin{align}
    \label{eq:qle-23}
    \mathbf{\dot{\hat{z}}}(t)&=\mathbf{A}\mathbf{\hat{z}}(t) + \mathbf{b}u(t)+\mathbf{\hat{k}}(t)[\zeta(t)-\mathbf{c}^{T}\mathbf{\hat{z}}(t)],\\
    \mathbf{\dot{\hat{\Sigma}}}(t)&=\mathbf{A}\mathbf{\hat{\Sigma}}(t)+\mathbf{\hat{\Sigma}}(t)\mathbf{A}^T+\mathbf{N}-[\mathbf{\hat{\Sigma}}(t)\mathbf{c}+\mathbf{M}][\mathbf{\hat{\Sigma}}(t)\mathbf{c}+\mathbf{M}]^{T},\label{eq:qle-23_riccati}\\
    \mathbf{\hat{k}}(t)&=\mathbf{\hat{\Sigma}}(t)\mathbf{c}+\mathbf{M},
  \end{align}
\end{subequations}
where $\zeta(t)\in \mathds{R}$ denotes the measurement signal resulting from a measurement of $y_{\lambda_0}^{\mathrm{out}}(t)$. These equations are correct for general Gaussian systems that can be
described by quantum Langevin equations of the form \eqref{eq:contStateSpace},
in particular also for systems where $\mathbf{M}\neq 0$. 
Note that although these equations are derived from a quantum description of the experiment, they are classical (stochastic) differential equations that involve classical quantities (the moments of $\mathbf{z}$ under $\hat{\rho}$, the measurement signal $\zeta$) only and can thus be readily implemented on a classical signal processor.

The results presented above show that the quantum filtering problem for Gaussian systems described by a quantum Langevin equation \eqref{eq:contStateSpace-z} (together with the output equation \eqref{eq:contStateSpace-y}) is formally equivalent to the classical filtering problem for the corresponding classical Langevin equation when using the correct noise properties \eqref{eq:qle-corr-matrices} that arise from a quantum description. For the details of the derivation in the framework of quantum filtering see \cite{hofer_chapter_2017}.
.

Additional to the approach taken in quantum filtering theory~\cite{bouten_introduction_2007}, complementary approaches exist to describe the dynamics of a (Gaussian) quantum system under continuous measurement.
These include a fully Gaussian treatment in a phase-space description \cite{genoni_conditional_2016} and the well-known quantum trajectories formalism \cite{Carmichael1993} which describes the stochastic evolution of the wave function.

\section{Optimal feedback cooling}\label{sec:kalman}

Online (optimal) estimation~\cite{kalman1960} and automatic control~\cite{kalman1960lqr,Stengel1994} techniques have become ubiquitous in modern technology~\cite{doyle2013feedback,aastrom2013computer, Moreno2009}. Due to the required level of control they are also becoming an increasingly important tool in quantum research and quantum technologies.

Here we design an optimal feedback controller in order to cool the particle's motion into the quantum ground state.
For linear (quantum) systems driven by Gaussian white noise, an optimal output feedback law can be obtained by solving the \ac{lqg} problem.
Its solution consists of the combination of a Kalman filter and a linear quadratic regulator, which can be designed independently of each other, as stated by the separation principle \cite{bouten_separation_2008}, breaking the design of the \ac{lqg} down into an estimation step and a control step.
The regulator computes the optimal feedback for a given state by solving an optimization problem in order to minimize the energy of the system.
Since the system state is in general not completely measurable, a Kalman filter is designed to provide optimal state estimates based on noisy measurements.
The basis of the design process of the \ac{lqg} is the mathematical description of the experimental setup detailed in the sections above. The experimental characterization of the involved quantities is described in detail in Section \ref{sec:parameteridentification}.

\subsection{Discretized time evolution}
While physical systems are usually considered in continuous time, estimation and control algorithms are necessarily implemented in a time-discrete manner. 
The resulting effects of the discretization process can be considered for linear dynamical systems by deriving a time-discrete formulation of the state-space model, evaluating it at times $t_k=k T_s$. 
To this end, we integrate \eqref{eq:contStateSpace} over a sampling time $T_s=t_{k+1}-t_k$ (which we assume is short on all system time scales), defining $\mathbf{z}_{k}{=}\mathbf{z}(t_k)$, $\mathbf{u}_{k}{=}\mathbf{u}(t_k)$, and the fundamental solution $\mathbf{\Phi}(t)=\exp(\mathbf{{A}}t)$. We find
\begin{equation}
    \begin{aligned}
    \mathbf{z}(t_{k+1})&=\mathbf{\Phi}(T_s)\mathbf{z}(t_{k}) + \int_{t_k}^{t_{k+1}}\dd{\tau}\mathbf{\Phi}(t_{k+1}-\tau)[\mathbf{b}u(\tau)+\mathbf{w}(\tau)]\\
    &= \mathbf{A}_{\mathrm{d}}\mathbf{z}(t_{k}) + \mathbf{b}_{\mathrm{d}}u(t_k)+\mathbf{\bar{w}}_k,
    \end{aligned}
\end{equation}
where we assumed that $u(t)$ is piecewise constant over the sampling time, i.e., $u(t) = u_k$ for $t\in[t_k,t_{k+1}]$ (zero-order hold used as a model for the digital-to-analog converter) and we introduced
the matrices $\mathbf{A}_{\mathrm{d}}=\exp\left(\mathbf{A} T_s\right)$ and $\mathbf{b}_{\mathrm{d}}=\int_0^{T_s} \exp\left(\mathbf{A}\tau\right)\mathbf{b}\mathrm{d}\tau$. The discretized noise process $\mathbf{\bar{w}}_k$ is given by $\mathbf{\bar{w}}_k=\int_{t_k}^{t_{k+1}}\dd{\tau}\mathbf{\Phi}(t_{k+1}-\tau)\mathbf{w}(\tau)$.

To describe the measurement, we define the time-averaged operator $\bar{y}^{\mathrm{out}}_{\lambda_0,k}{:=}\tfrac{1}{T_s}\int_{t_k}^{t_{k+1}}\dd{s}y^{\mathrm{out}}_{\lambda_0}(s)$ together with a corresponding expression for $\bar{y}_{\lambda_0,k}$. Assuming that $\mathbf{z}(t)$ likewise is approximately constant over the sampling time $T_s$ we find the discretized quantum state-space model
\begin{subequations}
\label{eq:disStateSpace}
    \begin{align}
        \mathbf{z}_{k+1}&=\mathbf{A}_{\mathrm{d}} \mathbf{z}_{k}+ \mathbf{b}_{\mathrm{d}}u_{k} + \mathbf{\bar{w}}_{k},\\
        \bar{y}^{\mathrm{out}}_{\lambda_0,k} &= \mathbf{c}^{\mathrm{T}}\mathbf{z}_{k} + \bar{y}_{\lambda_0,k}.\label{eq:noisyMeasurement}
    \end{align}
\end{subequations}
In analogy to \eqref{eq:qle-corr-matrices} the (cross-) correlations for
the noise processes $\mathbf{\bar{w}}_k$ and $\bar{y}_{\lambda_0,k}$ are given by
\begin{subequations}
\label{eq:discrete-corr-matrices}
  \begin{align}
    \mean{\bar{y}_{\lambda_0,k}\bar{y}_{\lambda_0,k'}}&=\bar{R}\delta_{kk'}=(1/T_s)\delta_{kk'},\\
    \mathrm{Re}\mean{\mathbf{\bar{w}}_k \bar{y}_{\lambda_0,k'}}&=\mathbf{\bar{M}}\delta_{kk'}=0,\\
    \mathrm{Re}\mean{\mathbf{\bar{w}}_k\mathbf{\bar{w}}^{T}_{k'}}&=\mathbf{\bar{N}}\delta_{kk'}\approx\mathbf{N}T_s\delta_{kk'},
  \end{align}
\end{subequations}
where the relation $\mathbf{\bar{N}}\approx\mathbf{N}T_s$ is true only if the sampling time is much shorter than all system time scales.

\subsection{Discrete-time Kalman Filter}
The Kalman filter for
the state-space system \eqref{eq:disStateSpace} is given by
\cite{edwards_duality_2003}
\begin{align}
             \label{eq:stateEstimator} \mathbf{\hat{z}}_{k+1}&= \mathbf{A}_{\mathrm{d}} \mathbf{\hat{z}}_{k} + \mathbf{b}_{\mathrm{d}} u_{k} + \mathbf{\hat{k}}\left( \zeta_{k} - \mathbf{c}^{\mathrm{T}} \mathbf{\hat{z}}_{k} \right), 
\end{align}
where $\zeta_k$ is the discretized measurement signal corresponding to
$\bar{y}^{\mathrm{out}}_{{\lambda_0},k}$ and the observer gain $\mathbf{\hat{k}}$ of the Kalman filter results from
\begin{align}\label{eq:kalmanGain}
  \mathbf{\hat{k}}&= \left(\mathbf{A}_{\mathrm{d}} \mathbf{\hat{\Sigma}^\mathrm{ss}}_{\mathrm{d}} \mathbf{c}+\mathbf{\bar{M}}\right) \left( \mathbf{c}^{\mathrm{T}} \mathbf{\hat{\Sigma}^\mathrm{ss}}_{\mathrm{d}} \mathbf{c}  + \bar{R}\right)^{-1}.
\end{align}
The (time-discrete) steady state error covariance matrix $\mathbf{\hat{\Sigma}^\mathrm{ss}}_{\mathrm{d}}$ is
computed by solving the discrete algebraic Riccati equation
\begin{align}\label{eq:kalmanRiccati}
  \mathbf{\hat{\Sigma}^\mathrm{ss}}_{\mathrm{d}}&=
  \mathbf{A}_{\mathrm{d}}\mathbf{\hat{\Sigma}^\mathrm{ss}}_{\mathrm{d}}\mathbf{A}_{\mathrm{d}}^{\mathrm{T}}
  + \mathbf{\bar{N}} - \left(\mathbf{A}_{\mathrm{d}} \mathbf{\hat{\Sigma}^\mathrm{ss}}_{\mathrm{d}}
  \mathbf{c}+\mathbf{\bar{M}}\right) \left(
  \mathbf{c}^{\mathrm{T}}\mathbf{\hat{\Sigma}^\mathrm{ss}}_{\mathrm{d}}\mathbf{c} +
  \bar{R}  \right)^{-1}\left(\mathbf{A}_{\mathrm{d}} \mathbf{\hat{\Sigma}^\mathrm{ss}}_{\mathrm{d}}
  \mathbf{c}+\mathbf{\bar{M}}\right)^{\mathrm{T}}.
\end{align}
Note that the Kalman filter \eqref{eq:stateEstimator} with the observer gain
\eqref{eq:kalmanGain} and the discrete algebraic Riccati equation
\eqref{eq:kalmanRiccati} is the time-discrete description of the Kalman--Bucy
filter \eqref{eq:kalmanBucy} and therefore describes the motional quantum state
of the nanosphere conditioned on the measurement, as shown in Section
\ref{sec:qlangevin}. In the limit of $T_s\rightarrow 0$ we recover the
Kalman--Bucy equations \eqref{eq:kalmanBucy} and $\mathbf{\hat{\Sigma}^\mathrm{ss}}_{\mathrm{d}}\rightarrow\mathbf{\hat{\Sigma}^\mathrm{ss}}$. As written, the Kalman filter is also valid for general systems with $\mathbf{\bar{M}}\neq 0$.

\subsection{Linear Quadratic Gaussian Regulator}
The concept of optimal feedback control consists of finding the optimal control inputs such that the system is stably operated at minimum cost. The optimal control input $u_{k}$ is obtained by minimizing the expected cost

\begin{align}\label{eq:costFunction}
    J\left( u_{k} \right) &=\lim\limits_{N \rightarrow \infty}\left\langle{\frac{1}{N}}
 \sum\limits_{k=0}^{N-1}\left(\mathbf{z}^{\mathrm{T}}_{k}\mathbf{Q} \mathbf{z}_{k}  + r u_{k}^{2} \right)\right\rangle
\end{align}
with respect to \eqref{eq:disStateSpace}, where $\langle\cdot\rangle$ refers to the quantum expectation value with respect to the initial state of the system and environment.
Here, the first term with weighting matrix
$\mathbf{Q}=\diag\left(\frac{\Omega_z}{2},\frac{\Omega_z}{2}\right)$
represents the total energy of the particle while the second term penalizes the required control effort scaled by
$r = \Omega_z/g_{\mathrm{fb}}^2$, with the feedback gain $g_{\mathrm{fb}}$ in units of \si{\radian\per\second}.
The control law that minimizes the cost function \eqref{eq:costFunction} is given by~\cite{edwards_duality_2003}
\begin{align}\label{eq:lqr}
    u_{k} &=-\mathbf{k}^{\mathrm{T}}{\mathbf{\hat{z}}}_{k}\text{~\@.}
\end{align}
The feedback vector $\mathbf{k}^{\mathrm{T}}$ is calculated by
\begin{align}
    \mathbf{k}^{\mathrm{T}}&= \left(r+ \mathbf{b}_{\mathrm{d}}^{\mathrm{T}}\mathbf{\Omega^\mathrm{ss}}\mathbf{b}_{\mathrm{d}}\right)^{-1} \mathbf{b}^{\mathrm{T}}\mathbf{\Omega^\mathrm{ss}}\mathbf{A}_{\mathrm{d}}
\end{align}
where $\mathbf{\Omega^{\mathrm{ss}}}$ is determined by the discrete algebraic Riccati equation
\begin{align}
  \mathbf{\Omega^\mathrm{ss}}&= \mathbf{Q}+\mathbf{A}_{\mathrm{d}}^{\mathrm{T}}\mathbf{\Omega^\mathrm{ss}}\mathbf{A}_{\mathrm{d}} - \mathbf{A}_{\mathrm{d}}^{\mathrm{T}}\mathbf{\Omega^\mathrm{ss}}\mathbf{b}_{\mathrm{d}}\left( r+\mathbf{b}_{\mathrm{d}}^{\mathrm{T}}\mathbf{\Omega^\mathrm{ss}}\mathbf{b}_{\mathrm{d}}  \right)^{-1}\mathbf{b}_{\mathrm{d}}^{\mathrm{T}}\mathbf{\Omega^{\mathrm{ss}}}\mathbf{A}_{\mathrm{d}}\text{~\@.}\label{eq:lqrRiccati}
\end{align}
The solution of the quantum LQG problem is thus formally identical to the one of the classical LQG problem for a classical state-space model of the form \eqref{eq:disStateSpace} and cost function of the form \eqref{eq:costFunction} (when interpreting $\langle \cdot \rangle$ as an appropriate classical expectation value).
In general, the observer gain $\mathbf{\hat{k}}$ and the feedback vector $\mathbf{k}^{\mathrm{T}}$ are time variant and they are calculated by solving the discrete Riccati equation for $\mathbf{\hat{\Sigma}}_{k}$ forwards in time and for $\mathbf{\Omega}_{k}$ backwards in time for a finite time horizon. If the time goes to infinity, the stationary solution $\mathbf{\hat{\Sigma}}_{k+1}= \mathbf{\hat{\Sigma}}_{k} = \mathbf{\hat{\Sigma}}^\mathrm{ss}_{\mathrm{d}}$ and $\mathbf{\Omega}_{k+1}=\mathbf{\Omega}_{k}=\mathbf{\Omega}^\mathrm{ss}$ of the corresponding discrete algebraic Riccati equation has to be calculated (see \eqref{eq:lqrRiccati}) and \eqref{eq:kalmanRiccati}). Thus, the \ac{lqg} becomes time invariant.
The transfer function of the time invariant \ac{lqg}, combining \eqref{eq:lqr} and the Kalman filter \eqref{eq:stateEstimator}, is given by
\begin{align}\label{eq:closedLoopTF}
    G(z)&=\frac{u_z(z)}{\zeta_z(z)}=-\mathbf{k}^{\mathrm{T}}\left(z \mathbf{I}-\left(\mathbf{A}_{\mathrm{d}}- \mathbf{b}_{\mathrm{d}}\mathbf{k}^{\mathrm{T}} - \mathbf{\hat{k}}\mathbf{c}^{\mathrm{T}} \right)\right)^{-1}\mathbf{\hat{k}}
\end{align}
where $u_z(z)$ and $\zeta_z(z)$ are the $\mathcal{Z}$-transform of the control input and measurement signal, $u_z(z)=\mathcal{Z}\left\{ \left( u_{k} \right)   \right\}$ and $\zeta_z(z)=\mathcal{Z}\left\{\left( \zeta_{k} \right) \right\}$, respectively, and $\mathbf{I}$ is the identity matrix.
The time discrete transfer function \eqref{eq:closedLoopTF} is implemented as a digital filter with a sampling time of $T_s=\SI{32}{\nano\second}$ on the \textsf{Red Pitaya} board which is equipped with a \textsf{Xilinx Zynq 7010} FPGA.

\begin{figure*}[!h]
    \includegraphics[scale=1]{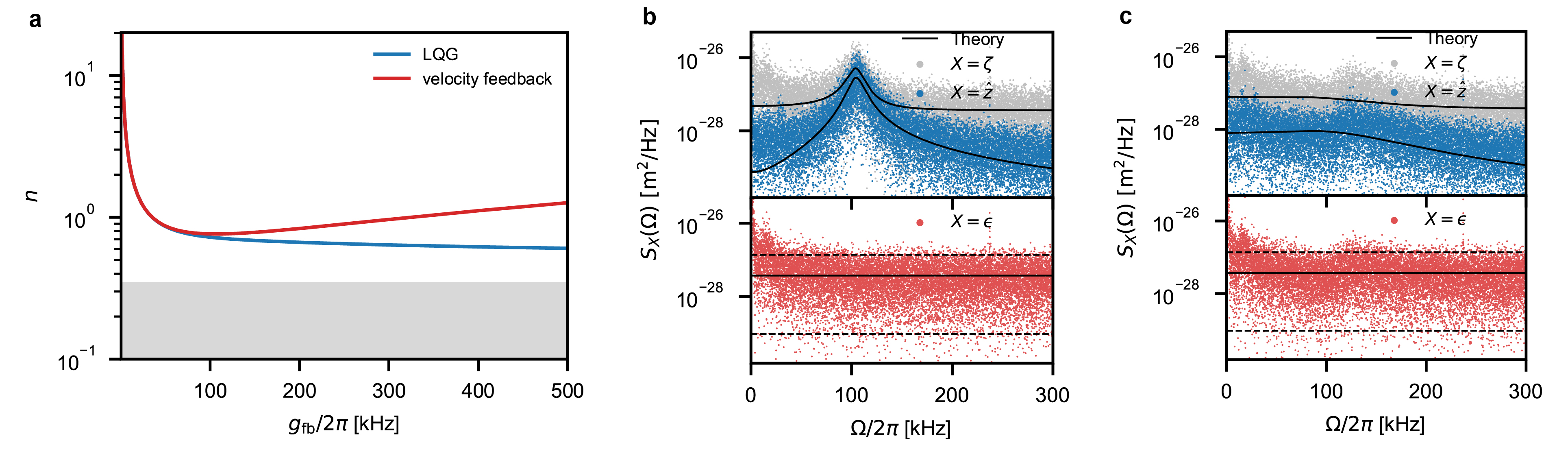}
    \caption{\textbf{LQG performance}. \textbf{a}, Comparison of the analytic solution of the occupation for optimal (LQG) and non-optimal (velocity) control and estimation methods. Despite the use of an optimal state estimator the closed-loop solution of the velocity feedback (red) is diverging for high feedback gains, contrary to the LQG (blue). This shows the importance of using the complete state vector in the feedback in order to minimize the energy of the system. \textbf{b - c}, Power spectral densities of the measurement (gray), Kalman estimation (blue), innovation (red) and the analytic solution of the mathematical description (black line) at $g_\mathrm{fb}/2\pi = \SI{16}{\kilo\hertz}$ and $g_\mathrm{fb}/2\pi = \SI{180}{\kilo\hertz}$. The black lines in the innovation plot indicate the white noise model (solid) and the $95\%$ confidence region of the expected $\chi^2$ distribution (dashed)~\cite{Wieczorek2015}.}
    \label{fig:SI_LQG1}
\end{figure*}
The effects of the low frequency $1/f$ phase noise and the intrinsic delay of the controller of about $300 \mathrm{ns}$ are negligible in a fairly large frequency band around resonance, and at most of the feedback gains we are operating at. For this reason we do not include these effects into the model, in favor of a larger dynamic range for the output.
We observe, however, a drift in the oscillation frequency for increasing feedback gains which is caused presumably by nonlinear effects not captured by the mathematical model. This error leads to the appearance of color in the innovation sequence and a decreasing cooling performance. The calibration of the measurement signal and feedback force, as well as the characterization of the noise processes $\bar{y}_{\lambda_0,k}$ and $\bar{w}_{k}$ are presented in the following section.

\subsection{Colored Noise Model}\label{sec:pinkman}
\begin{figure*}[!h]
    \includegraphics[scale=1]{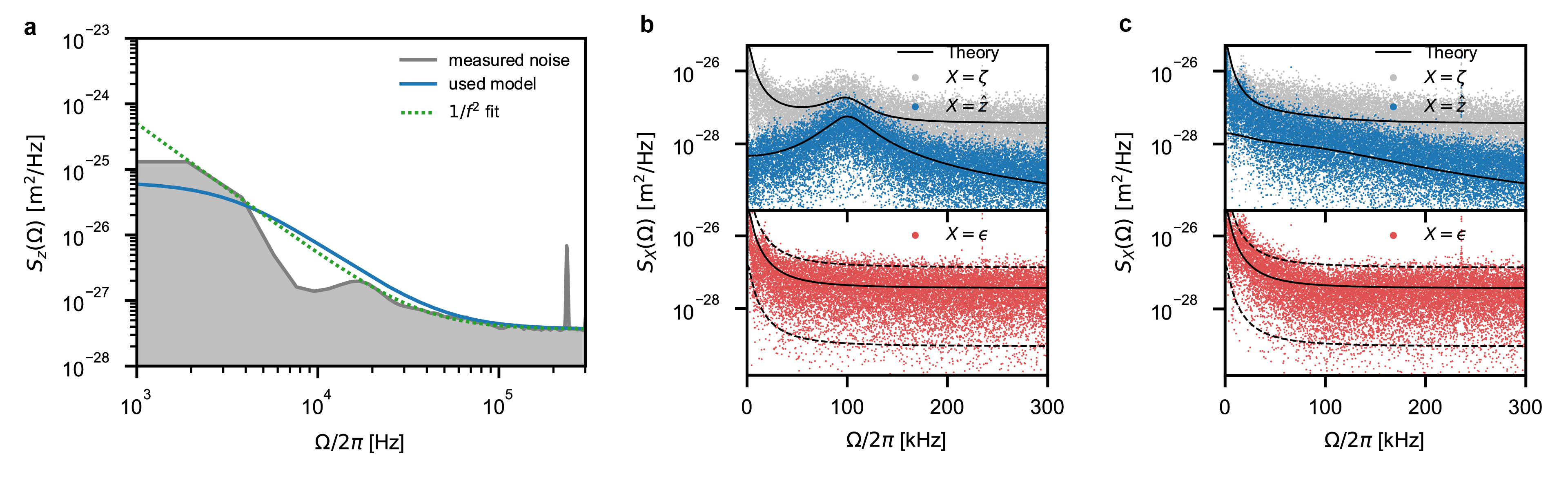}
    \caption{\textbf{Colored noise model}. \textbf{a} Comparison of the power spectral densities of the Brownian noise model (green) and the low-pass noise model (blue) the power spectral densities of the innovation (gray). \textbf{b - c}, Power spectral densities of the measurement (gray), Kalman estimation (blue), innovation (red) and the analytic solution of the mathematical description (black line) at $g_\mathrm{fb}/2\pi = \SI{40}{\kilo\hertz}$ and $g_\mathrm{fb}/2\pi = \SI{150}{\kilo\hertz}$. The extension with an appropriate noise model brings along more accurate estimates of the state. The black lines in the innovation plot indicate the colored noise model (solid) and the $95\%$ confidence region of the expected $\chi^2$ distribution (dashed)~\cite{Wieczorek2015}.}
    \label{fig:SI_LQG2}
\end{figure*}

Although the effects of low frequency noise are negligible compared to the white noise level, we have seen that this model mismatch is amplified by the controller, and would eventually be limiting the closed-loop performance at feedback gains larger than $200\mathrm{kHz}$. For this reason, we also extend the state-space model (\ref{eq:contStateSpace}) by an appropriate colored noise model. The Kalman filter is designed on the basis of an  extended state-space model given by:
\begin{subequations}\label{eq:pinkmanStateSpace}
    \begin{align}
        \dot{\tilde{\mathbf{z}}}(t)&=\tilde{\mathbf{A}} \tilde{\mathbf{z}}(t) + \tilde{\mathbf{b}} u(t) + \tilde{\mathbf{G}} \tilde{\mathbf{w}}(t)\,, & \qquad    \tilde{\mathbf{x}}(0)=\tilde{\mathbf{x}}_0 \\
        y^{\mathrm{out}}_{\lambda_0}(t) &= \tilde{\mathbf{c}}^{\mathrm{T}} \tilde{\mathbf{z}}(t) + y_{\lambda_0}(t)
    \end{align}
\end{subequations}
with the extended state vector $\tilde{\mathbf{z}}(t)=\begin{bmatrix} \mathbf{z}(t)^{\mathrm{T}}& \mathbf{\xi}(t)^{\mathrm{T}}\end{bmatrix}^{\mathrm{T}}$, and the process noise input vector $\tilde{\mathbf{w}}(t)=\begin{bmatrix}w(t) & \mu (t)  \end{bmatrix}^{\mathrm{T}}$, where $\mu (t)$ is white Gaussian noise, which drives the chosen noise model.
The extended system matrix $\tilde{\mathbf{A}}$, the extended input vector of the control input $\tilde{\mathbf{b}}$, the extended input matrix of the process noise $\tilde{\mathbf{G}}$ and the extended output vector $\tilde{\mathbf{c}}^{\mathrm{T}}$ are defined as
\begin{align*}
    \tilde{\mathbf{A}}&=\begin{bmatrix}
       \mathbf{A}&\mathbf{0}\\\mathbf{0}&\mathbf{A}_n
    \end{bmatrix},
    &\tilde{\mathbf{b}}&=\begin{bmatrix}
        \mathbf{b}\\\mathbf{0}
    \end{bmatrix},
    &\tilde{\mathbf{G}}&=\begin{bmatrix}
        \mathbf{g}&\mathbf{0}\\\mathbf{0}& \mathbf{g}_{n}
    \end{bmatrix},
    &\tilde{\mathbf{c}}^{\mathrm{T}}&=\begin{bmatrix}
        \mathbf{c}^{\mathrm{T}}&\mathbf{c}_{n}^{\mathrm{T}}
    \end{bmatrix},
\end{align*}
with the dynamic matrix of the noise model $\mathbf{A}_{n}$, the input vector of the noise model $\mathbf{g}_{n}$ and the output vector of the noise model $\mathbf{c}_{n}^{\mathrm{T}}$.
As proposed in \cite{Wieczorek2015}, Brownian noise is a good approximation for the non-white amplitude and phase noise of a laser, which is modeled by the state-space system \eqref{eq:pinkmanStateSpace} with $\mathbf{A}_{n}=0$ and $\mathbf{g}_{n}=\mathbf{c}_{n}^{\mathrm{T}}=1$. This model (green line in Figure \ref{fig:SI_LQG2}\textbf{a}) provides a good approximation of the low frequency noise that we observe. Nevertheless, it has a limited hardware feasibility as the magnitude of the noise becomes large in the lower frequency range and the slow dynamics lead to numerical issues in the fixed-point implementation, resulting in drift and overflows, and can even destabilize the closed-loop system.
For a practical hardware implementation we model the noise as a low-pass filter driven by white noise $\mu(t)$.
Thereby,  $\mathbf{A}_{n}$, $\mathbf{g}_{n}$ and  $\mathbf{c}_{n}^{\mathrm{T}}$ are obtained from the state-space representation of the low-pass filter $ G_{lp}\left( s \right)=1/\left( 1+s/\omega_{c} \right)$, with the cutoff frequency $\omega_{c}=\SI{3.5}{\kilo\hertz}$.
In Figure \ref{fig:SI_LQG2}\textbf{b}), we show the power spectral densities of the measurement (gray), the Kalman estimation (blue), the innovation (orange) in good agreement with the analytic solution of the mathematical description (black line) of the \ac{lqg} based on the proposed low-pass noise model for $g_\mathrm{fb}/2\pi=\SI{40}{\kilo\hertz}$. Nevertheless, for high gain feedback  $g_\mathrm{fb}/2\pi=\SI{150}{\kilo\hertz}$ (Figure \ref{fig:SI_LQG2}c)) the performance decreases significantly due to the reduced dynamic range of the hardware implementation to fit the more complex filter on the FPGA.
The use of a more powerful hardware would overcome such implementation issues of the Kalman filter using colored noise models and has the potential to further increase the performance.

\subsection{FPGA implementation}\label{sec:fpga}
The designed \ac{lqg} is implemented on a \textsf{Red Pitaya} board equipped with a \textsf{Xilinx Zynq 7010} FPGA.
The base design of the \textsf{Vivado Design Suite} project of the \textsf{Red Pitaya} is based on the tutorial provided by Anton Poto\v{c}nik \cite{Redpitaya}, modified to suit our purposes.
The time-invariant transfer function of the \ac{lqg} \eqref{eq:closedLoopTF} is implemented in \textsc{Matlab}/\textsc{Simulink} as digital filter with the \textsf{Xilinx System Generator for DSP}.
Thereby, hardware-in-the-loop simulations can be performed in \textsc{Matlab}/\textsc{Simulink}, capturing the exact behavior of the real implementation on the FPGA of the \textsf{Red Pitaya}.
This provides the possibility to quickly identify and fix issues with the fixed-point arithmetic and quantization.
The \textsf{Xilinx System Generator for DSP} allows automatic code generation of the designed filter, considering the ressource limitations and timing constraints of the FPGA.
The obtained VHDL code (IP Core) is inserted in the base design of the FPGA in the \textsf{Vivado Design Suite} and the bitstream file of the FPGA is generated.
Parameters can be changed online via communication with the AXI-bus.
The low frequency output noise of the \textsf{Red Pitaya} has been improved by removing the 2 resistors and disconnecting the noisy output offset~\cite{Redpitaya_hack}.

\section{Identification of the model parameters}
\label{sec:parameteridentification}
The identification of the system parameters is crucial for a properly tuned model based Kalman filter and \ac{lqr} design. Direct measurement of most of the system parameters depends on a proper calibration of the measurements. 

\subsection{Calibration of the measurement transduction coefficient}
In this section, we measure the calibration factor ($C_\mathrm{mV} \, [\mathrm{m/V}]$) for our homodyne detection. One possibility is thermometry in an environment in which the nanoparticle thermalizes to a room-temperature gas~\citep{Hebestreit2018}. Given the high resolution of our position measurement, the limited dynamic range of our detector and data acquisition board this method cannot be implemented directly, but would require multiple steps of amplification. In addition, the accuracy of this approach was verified only up to a factor of 2~\citep{Tebbenjohanns2020}.
To reconstruct the relation between the displacement of the particle in  meters and the homodyne time traces in volts we take advantage of the simultaneous out-of-loop measurement of the particle's energy via Raman thermometry (see Section \ref{sec:heterodyne}) at different feedback gains. To minimize the effects of noise squashing~\citep{Poggio2007} due to imprecision noise driving the motion of the particle via the feedback, we restrict calibration to low values of the feedback gain. We perform the calibration in an iterative way, where we alternate evaluation of the calibration factor and update of the controller setting.
The variance of the particles motion can be estimated from a measurement of energy in units of motional quanta $\left\langle n \right\rangle$ by $ \left\langle z^2 \right\rangle = z_\mathrm{zpf}^2 (2\left\langle n \right\rangle+1)$.
We compare this value with the variance of the signal $V(t)$ obtained from the homodyne noisy position measurement:
\begin{equation}
      \left\langle V^2 \right\rangle = C_\mathrm{mV}^{-2}\left(\left\langle z^2 \right\rangle + \left\langle \nu^2 \right\rangle\right) =  \int_{0}^{+\infty}S_{\zeta}(\Omega)\frac{d\Omega}{2\pi}.
\end{equation}
Where $\nu(t)$ is the measurement noise and $C_{_\mathrm{mV}}$ the calibration factor converting the measured voltage into the corresponding displacement in meters.
\begin{figure*}[!h]
\includegraphics[scale=1]{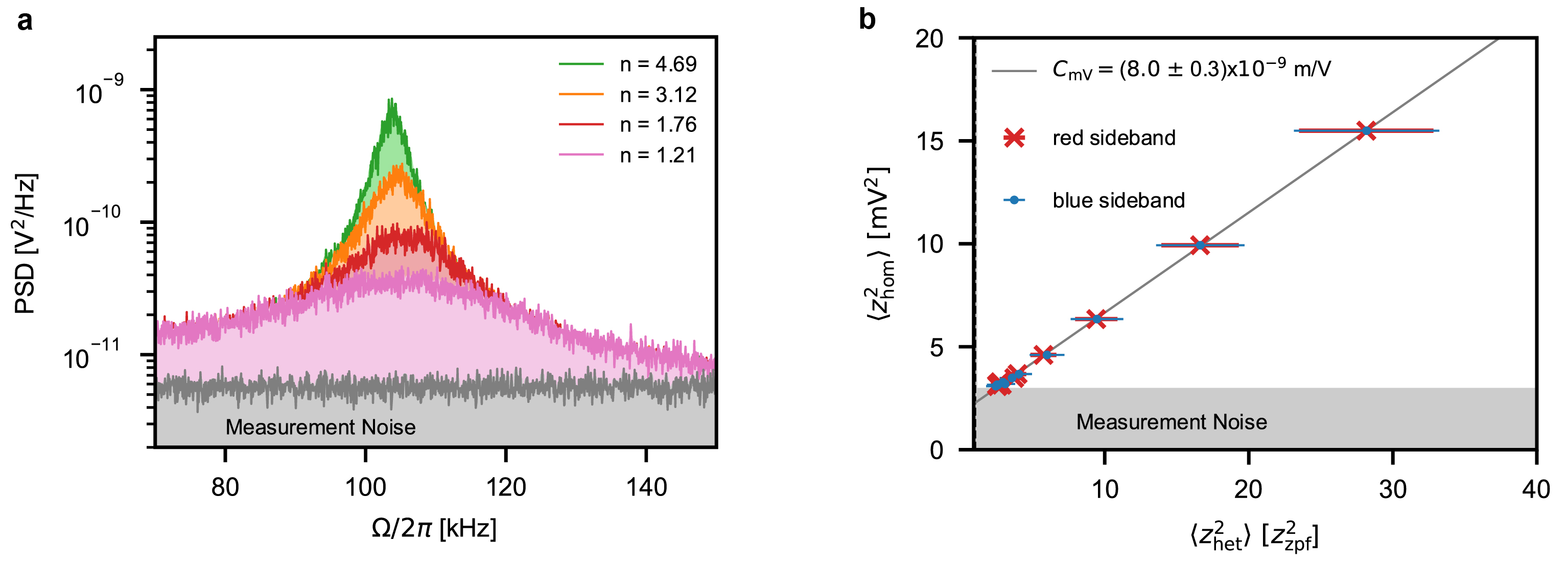}
\caption{\textbf{Position calibration}. \textbf{a} Measurement of the displacement power spectral densities at different feedback gains and labelled by the occupation measured independently by Raman thermometry. \textbf{b} Integrated voltage variance from the homodyne measurement plotted as a function of position variance extimated from the heterodyne measurement. Red crosses and blue dots represent the position variance estimated by the Stokes and anti-Stokes sidebands respectively. A linear fit provides the calibration factor for the homodyne measurement.}
\end{figure*}
We fit a linear function, where the offset indicates the measurement noise and the slope determines the calibration factor:
\begin{equation}
 C_\mathrm{mV} = (8.0\pm 0.3)\times10^{-9} \,\mathrm{m/V}
\end{equation}
We also verify the consistency of the measured calibration factor by considering all transduction coefficients composing the measurement. 
The phase-shift induced by the particle's displacement on to the fraction of collected light defines the measurement strength of our detection (what in cavity-optomechanics you would call $2G/\kappa$~\citep{Aspelmeyer2014}):
\begin{equation}
\chi = \frac{\partial \varphi}{\partial z} =\sqrt{\frac{\eta_{\mathrm{d,c}}}{\eta^*_{\mathrm{d,c}}}} \sqrt{\left(A^2+\frac{2}{5}\right)} k \quad \mathrm{[rad/m]}    
\end{equation}
In a homodyne detection the signal light beam is interfered with a strong local oscillator, and phase shifts are transduced to a power variation by $G_\mathrm{HOM}=2\sqrt{P_\mathrm{S} P_\mathrm{LO}}\,\mathrm{[W/rad]}$, where $P_\mathrm{S}=P_\mathrm{scatt}\eta^*/\eta^*_\mathrm{d,q}$ is the signal light just before the detector, $P_\mathrm{LO}$ is the local oscillator power, $\eta_\mathrm{d}^*$ the photon detection efficiency and $\eta_\mathrm{d,q}^*$ the detector quantum efficiency as defined in Section \ref{sec:losses}. Optical power is converted into an electron current at the photodiodes via  the detector responsivity is $R_\mathrm{det} = -e\eta_\mathrm{d,q}/\hbar\omega_0\,\mathrm{[A/W]}$, and finally the transimpedance gain $g_\mathrm{t} = 250\times10^3\,\mathrm{[V/A]}$ converts this current to a voltage. Impedance matching to the detector's $\SI{50}{\ohm}$ output attenuates the signal by $\SI{3}{\decibel}$. 
We can now convert measured voltage to meters by:
\begin{equation}
    C_\mathrm{mV} = \frac{\hbar\omega_0}{(-e)\eta_\mathrm{d,q} g_\mathrm{t} \sqrt{P_\mathrm{S}P_\mathrm{LO}} \chi} = 7.8\times10^{-9} \, \mathrm{m/V}
\end{equation}
which is in good agreement with the measured value. 

\subsection{Evaluation of the measurement noise}

From the calibrated PSD, we can measure the measurement noise level at the relevant frequencies. While the detector bandwidth is about 75~MHz, the Red Pitaya has a measurement bandwidth of 31.25~MHz. We include an anti-aliasing analog low pass filter with cut off at 11~MHz, below the sampling Nyquist frequency. This allows to minimize the aliasing of high frequency noise at the relevant frequencies. We measure the imprecision noise $S_{z}^\mathrm{imp}$ dominated by photon shot noise of the local oscillator by covering the signal beam. This results in a variance of measurement noise (assuming a white noise model) of
\begin{equation}
     \sigma_z^2 = S_{z}^\mathrm{imp} \frac{f_s}{2} = ( 5.4\pm 0.2 )\times10^{-21}\,\mathrm{m^2}
    \label{eq:measnoise}
\end{equation}
where $f_s =1/T_s = 31.25~\mathrm{MHz}$ is Red-Pitaya sampling frequency.
The measurement noise can likewise be estimated by evaluating the signal variance from independently characterized experimental parameters. This includes contributions of photon shot noise and detector dark noise:
\begin{equation}
   \left\langle V^2 \right\rangle = \left[\left(\frac{g_\mathrm{t}}{2} e \sqrt{\frac{P_\mathrm{LO}\eta_\mathrm{q}}{\hbar\omega_0}}\right)^2+\left(\frac{g_\mathrm{t}}{2}NEC \right)^2\right]f_s
\end{equation}
where the factor 2 below $g_\mathrm{t}$ arises from the coupling of the detector to $\SI{50}{\ohm}$ load and $NEC$ is the noise equivalent current. The noise equivalent position variance is therefore:
\begin{equation}
     \sigma_z^2   = C_\mathrm{mV}^2 \left\langle V^2 \right\rangle = 5.3 \times10^{-21}\,\mathrm{m^2}
\end{equation}
in good agreement with the measured value.

\subsection{Calibration of the applied force}

With the position calibration at hand, we can further map the applied voltage to the control electrode on the force acting on the charged nanoball. We drive the particle by applying a sinusoidal signal of known amplitude and frequency. In the case of strong off-resonant drive force $F_\mathrm{d}(t)$, with spectral density $S_{F}^\mathrm{d}$, if at a particular drive frequency $\Omega_\mathrm{d}$, having $S^\mathrm{d}_{F}(\Omega_\mathrm{d})\gg S^\mathrm{tot}_{F}(\Omega_\mathrm{d})$, the driven motion is related to the drive by: 
\begin{equation}
    S_{z}(\Omega_d) = S^{v}_{F}(\Omega_\mathrm{d})\lvert \mathrm{m}(\Omega_\mathrm{d})\rvert^2 + S_{z}^{\mathrm{imp}} 
\end{equation}
which in the simple case of $F_\mathrm{d}(t) = F_{\mathrm{d}0} \sin{(\Omega_\mathrm{d} t)} $, and $\Omega_\mathrm{d}/\gamma \gg 1$ results in:
\begin{equation}
    \left\langle z_\mathrm{d}^2 \right\rangle = \frac{1}{2 \pi} \int_{\Omega_\mathrm{d}-\epsilon}^{\Omega_d+\epsilon}\left(S_{z}(\Omega) - S_{z}^{\mathrm{imp}} \right)\, d\Omega =  \frac{\left\langle F_\mathrm{d}^2 \right\rangle}{(m (\Omega_z^2 - \Omega_{\mathrm{d}}^2))^2} = \frac{F_{\mathrm{d}0}^2/2}{(m (\Omega_z^2 - \Omega_{\mathrm{d}}^2))^2}.
\label{eq:calibration}
\end{equation}
The variance of the displacement is again obtained by integrating over the symmetrized spectral density around the driving frequency subtracting the background imprecision noise. 
As the driving force is proportional to the applied voltage $F_\mathrm{d}(t) = C_\mathrm{NV} V (t)$, we can use the relation \eqref{eq:calibration} to calibrate this to the applied force in newton and identify the transduction coefficient $C_\mathrm{NV}$. 
We perform the measurement at different values of drive frequency and amplitude, and plot the standard deviation of the calibrated force $\sqrt{ \left\langle F_v^2 \right\rangle} = m\sqrt{ \left\langle  z^2 \right\rangle (\Omega_z^2 - \Omega_\mathrm{d}^2)}$ versus the standard deviation of the applied signal. The slope gives a factor of $C_\mathrm{NV} = (1.98\pm 0.06)\times 10^{-15}\,\mathrm{N/V}$. The measurements at the two different frequencies result in perfectly overlapping values. 
\begin{figure*}[!h]
\includegraphics[scale=1]{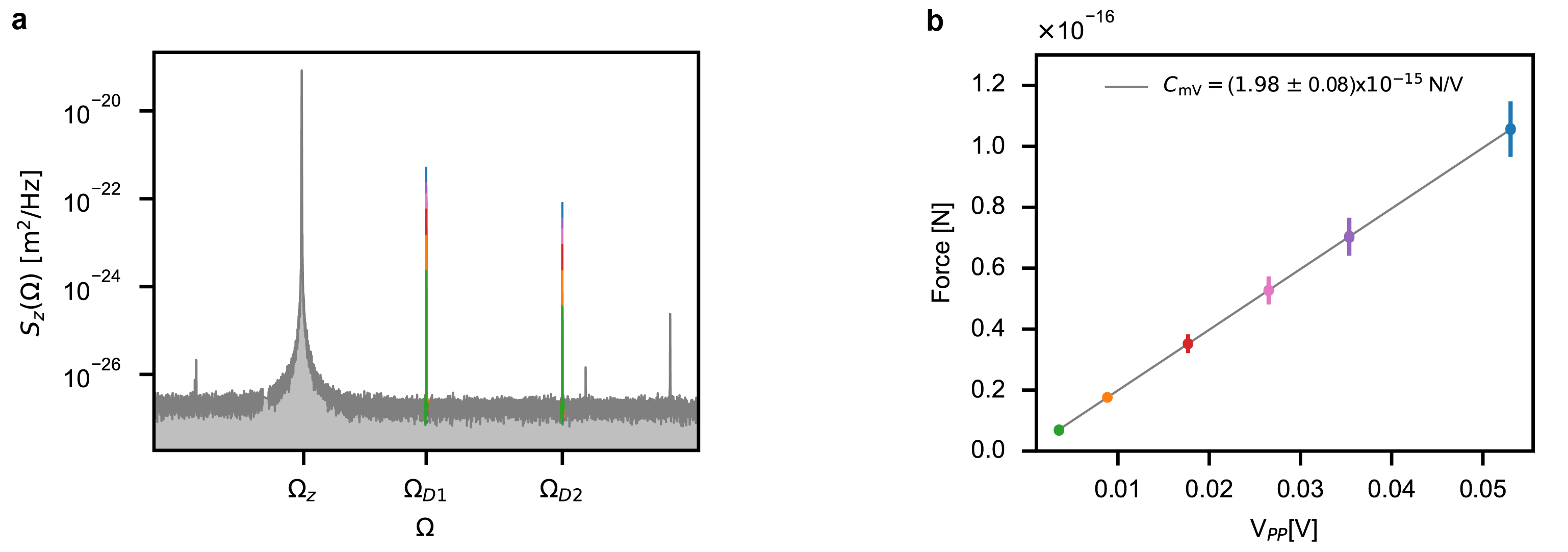}
\caption{\textbf{Force calibration}. To map the applied voltage [V] to a force [N], we drive the particle with a series of sinusoidal signals of different amplitude and frequency and measure the particles response in the calibrated position PSD.}
\end{figure*}

\subsection{Measurement of the thermal and backaction decoherence rates} 

\begin{figure*}[!h]
\includegraphics[scale=1]{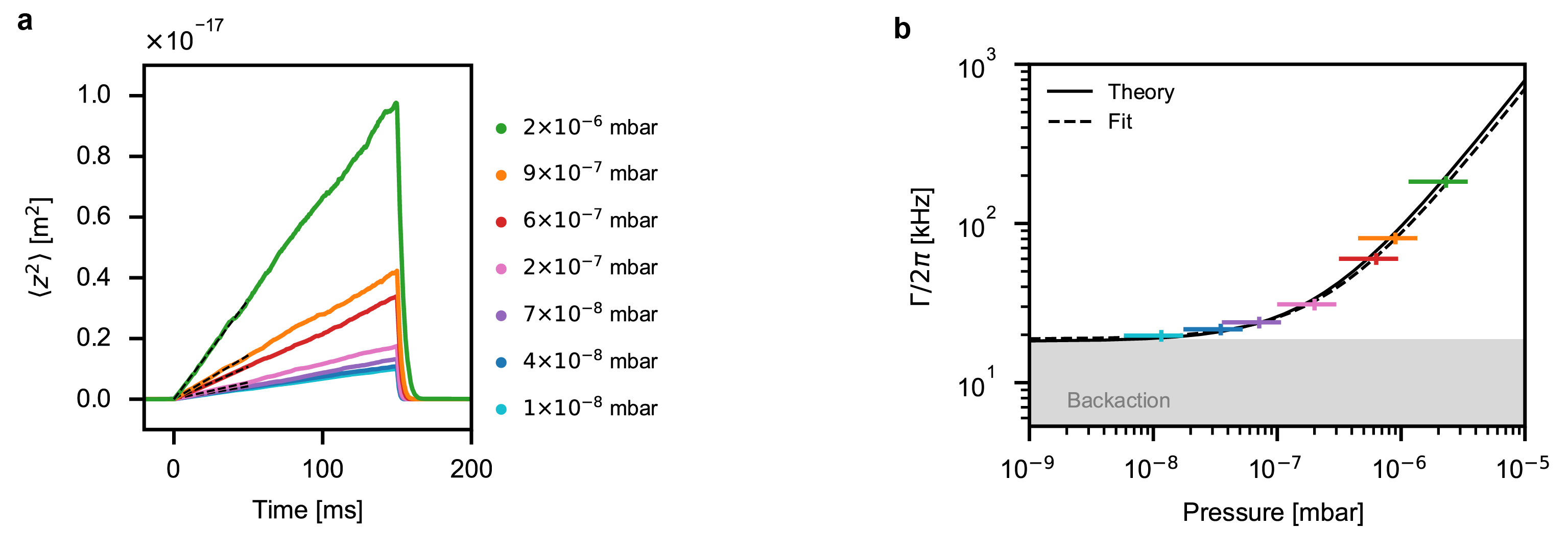}
\caption{\textbf{Heating rate}. The backaction and thermal contribution to the force noise are directly measured by performing re-heating measurements. We restrict the measurement to short ($150\,\mathrm{ms}$) re-heating periods. Longer ring up measurements may lead to the loss of the particle when mainly coupled to the high temperature photon bath. \textbf{a} At each pressure we release the feedback and observe the heating dynamics of the particle 1000 times. The ensemble average of the variance of these traces represents the average energy increase rate. \textbf{b} Pressure dependence of the heating rate. At pressures below $1\times 10^{-8}\, \mathrm{mbar}$ the contribution to the total force noise is dominated by the photon recoil, or measurement backaction. Horizontal error bars are given by the 50\% accuracy specified by the pressure gauge producer.}
\label{fig:heating}
\end{figure*}

We define the decoherence rates originating from thermal force noise and measurement backaction as the average rate of phonons delivered to the particle.
To determine the decoherence rate induced by measurement backaction and interactions with the thermal environment, we perform a set of re-heating measurements of the particle's energy. We do so by switching off the feedback, observing the relaxation trace. Ensemble averaging over many cycles allows to extract the average heating rates~\cite{Gieseler2014,Jain2016} (Figure \ref{fig:heating}a). To distinguish contributions from photon recoil (backaction) and gas collisions (thermal force), this is done at various pressures (Figure \ref{fig:heating}b).
When switching the feedback off, the energy $E$, or level of excitation of the oscillator $n = E/\hbar \Omega_z$ increases on average as:
\begin{equation}
    n(t) = n_{0} + n_\mathrm{th}(1-e^{-\gamma_\mathrm{th} t}) + n_\mathrm{ba}(1-e^{-\gamma_\mathrm{ba} t}) \overset{t\ll \frac{1}{\gamma_\mathrm{th,ba}}}{\approx} n_{0} + n_\mathrm{th}\gamma_\mathrm{th} t + n_{ba}\gamma_\mathrm{ba} t = n_{0} + (\Gamma_\mathrm{th} + \Gamma_\mathrm{ba}) t,
\label{eq:heating0}
\end{equation}
where $n_\mathrm{0}$ is the initial occupation and $n_\mathrm{th}$ and $n_\mathrm{ba}$ are the occupations associated to the thermal and optical baths respectively, $\gamma_\mathrm{th}$ is the gas damping and $\gamma_\mathrm{ba}$ the radiation damping that results from relativistic effects~\cite{Novotny2017}.
The decoherence rates are now written as:
\begin{equation}
\Gamma_\mathrm{th} =\gamma_\mathrm{th} n_\mathrm{th}
\quad
\mathrm{and}
\quad
\Gamma_\mathrm{ba} = \gamma_\mathrm{ba} n_\mathrm{ba},
\label{eq:heating}
\end{equation}

The thermal heating rate is derived by considering a thermal bath of energy $E_\mathrm{th} = k_\mathrm{B}T$ and temperature of $\SI{292}{K}$ and a damping rate given by \cite{Beresnev1990}:
\begin{equation}
    \gamma_\mathrm{th}= \frac{6\pi \eta_\mathrm{v} r}{m}\left(\frac{Kn}{0.619+Kn}\right)\left(1+\frac{0.310 Kn}{Kn^2+1.152 Kn+0.785}\right),
\label{eq:damping0}
\end{equation}{}
where $\eta_\mathrm{v}$ is the dilute gas shear viscosity, $Kn = \lambda_\mathrm{gas}(P)/L$ the Knudsen number, $\lambda_{gas}(P)$ the pressure dependent molecule mean free path, $L = V/A = 4r/3 $ the particle's characteristic length, $V$ its volume and $A$ its cross section. In the low pressure limit, $Kn\gg 1$, eq.~\eqref{eq:damping0} can be approximated by:
\begin{equation}
    \gamma_\mathrm{th} =\frac{64}{3} \frac{r^2 P}{m \bar{v}_\mathrm{gas}},
\label{eq:damping}
\end{equation}
where $\bar{v}_\mathrm{gas} = \sqrt{8 R T/(\pi m_\mathrm{gas})}$ is the mean gas velocity, $r$ and $m$ the particle radius and mass respectively, $P$ the pressure (expressed in Pascal, SI), $R = k_\mathrm{B} N_\mathrm{A} = 8,3144\ \mathrm{J/(mol\ K)}$ the universal gas constant, and $m_\mathrm{gas}$ the molar mass of the gas. Up to $\SI{1}{mbar}$, this approximation exhibits a deviation of less than $10^{-2}$ from the real value for the particles we are considering.
Following the treatment by Seberson and Robicheaux~\cite{Seberson2019}, we can also derive the the contribution of photon recoil to the heating reate:
\begin{equation}
    \Gamma_{\mathrm{ba}, z} = \left(A^2+\frac{2}{5} \right)\frac{\omega_0 P_\mathrm{scatt}}{2 \Omega_z m c^2 }
    \label{eq:heatingrateba}
    \end{equation}

At a pressure of $1.6 \times 10^{-8}\,\mathrm{mbar}$ we directly measure a minimal total heating rate of:
\begin{equation}
    \Gamma_\mathrm{tot} = \Gamma_\mathrm{th} +\Gamma_\mathrm{ba} = 2\pi \cdot(19.7 \pm 1.5 ) \,  \mathrm{kHz}
\end{equation}
With a linear fit to the pressure dependent data (Figure \ref{fig:heating}) we can extrapolate the contributions of thermal noise and measurement backaction at all pressures, finding them in excellent agreement with the values estimated using equations \ref{eq:damping} and \ref{eq:heatingrateba} in our experimental settings (Figure \ref{fig:heating}b).
At the minimal operating pressure $9.2 \times 10^{-9}\,\mathrm{mbar}$, we find the process noise to be (for the Kalman filter)
\begin{equation}
    \sigma_F^2 = \left\langle F_\mathrm{tot}^2 \right\rangle =  S_{F}^\mathrm{tot} \frac{f_s}{2} = 4\hbar\Omega_z m\Gamma_\mathrm{tot} \frac{f_s}{2} = (1.5 \pm 0.1) \times 10^{-33} \,  \mathrm{N}^2
\label{eq:forcenoise}
\end{equation}
 
\section{Raman scattering thermometry}
\label{sec:heterodyne}

The optomechanical interaction exhibits both energy and momentum exchange between the oscillating particle and the elecromagnetic filed. While elastic scattering (Rayleigh) leave the energy of the scattered photons unaltered, the side-bands of $n^\mathrm{th}$ order in the absorption and fluorescence spectra due to inelastic scattering (Raman) are interpreted as transitions between the quantized energy levels of the harmonic oscillator~\cite{Jessen1996}. The elastic and inelastic scattering rates can be calculated using Fermi's golden rule:
\begin{equation}
    \Gamma_{n\rightarrow n+\Delta n } =\frac{2\pi}{\hbar} M_{n, n+\Delta n} \rho(n+\Delta n)
\end{equation}
where $\rho(n+\Delta n)$ the population density of occupation $n+\Delta n$ state of the particle motion and $M_{n, n+\Delta n}$ the transition matrix element given by the  cross term in the dipole-field interaction~\cite{Oriol2011,Sinha2020}: 
\begin{equation}
    M_{n, n+\Delta n} =\lvert\bra{n+\Delta n}\hat{H}_\mathrm{I}\ket{n}\rvert^2  \propto \lvert\bra{n+\Delta n}(\chi z_\mathrm{zpf}(b+b^{\dagger}))^{\Delta n}\ket{n}\rvert^2= (\chi z_\mathrm{zpf})^{2\Delta n}\text{~\@.}
\end{equation}
Here $\hat{H}_\mathrm{I} \propto \hat{a} e^{-i\chi \hat{z}}+\mathrm{H.c.}$ with $\hat{z} = z_\mathrm{zpf}(\hat{b}+\hat{b}^{\dagger})$, and $\chi$ the mean momentum transferred to the particle by a photon scattered into the detection mode~\cite{Jessen1996}.
As the  transition matrix element is symmetric, the asymmetry of the scattering rates into the Stokes and anti-Stokes sideband arises from population differences between the vibrational states. Moreover, considering a thermal steady state, the ratio between first order ($\Delta n =1$) transition rates is given by the detailed balance $\Gamma_{\mathrm{S}} \rho(n)= \Gamma_{\mathrm{aS}}\rho(n+1)$~\cite{Clerk2010}: 
\begin{equation}
\frac{\Gamma_{\mathrm{aS}}}{\Gamma_{\mathrm{S}}}=\frac{\Gamma_{n + 1 \rightarrow n}}{\Gamma_{n\rightarrow n+1 }} = \frac{\rho(n)}{\rho(n+1)} = e^{\frac{\hbar\Omega_z}{k_\mathrm{B} T}}   = R\text{~\@.}
\label{eq:scatteringrate}
\end{equation}
From this ratio one can extract the average occupation for a thermal state defined as:
\begin{equation}
    \langle n \rangle = \frac{1}{e^{\frac{\hbar\Omega_z}{k_\mathrm{B} T}}-1} = \frac{R}{R-1}
\end{equation}
In absence of a cavity, the motion of the mechanical oscillator interacts with a white continuum vacuum state, and the mechanical power spectral density is linearly transduced to the output optical state. The measured heterodyne optical power spectral density describes the ability of the optical field to absorb (yield) energy from (to) the mechanical oscillator~\cite{Weinstein2014}. The first order power spectral density for the quantum harmonic oscillator is~\cite{Clerk2010,Hauer2015}:
\begin{equation}
S_{zz}(\Omega) = z_\mathrm{zpf}^2 \gamma \left[ \frac{n+1}{(\omega + \Omega_z)^2 + (\gamma/2)^2}  +
 \frac{n}{(\omega - \Omega_z)^2 + (\gamma/2)^2} \right].
\label{eq:heterodyne}
\end{equation}
The scattering rates of the two competing processes ($\Gamma_{\mathrm{S}}$, $\Gamma_{\mathrm{aS}})$ correspond to the powers detected in the sidebands of the heterodyne measurement (Figure \ref{fig:hetfit}), allowing from such a measurement, direct evaluation of the motional energy of the thermal harmonic oscillator.

\subsection{Heterodyne noise analysis}
\begin{figure*}[!htb]
\includegraphics[scale=1]{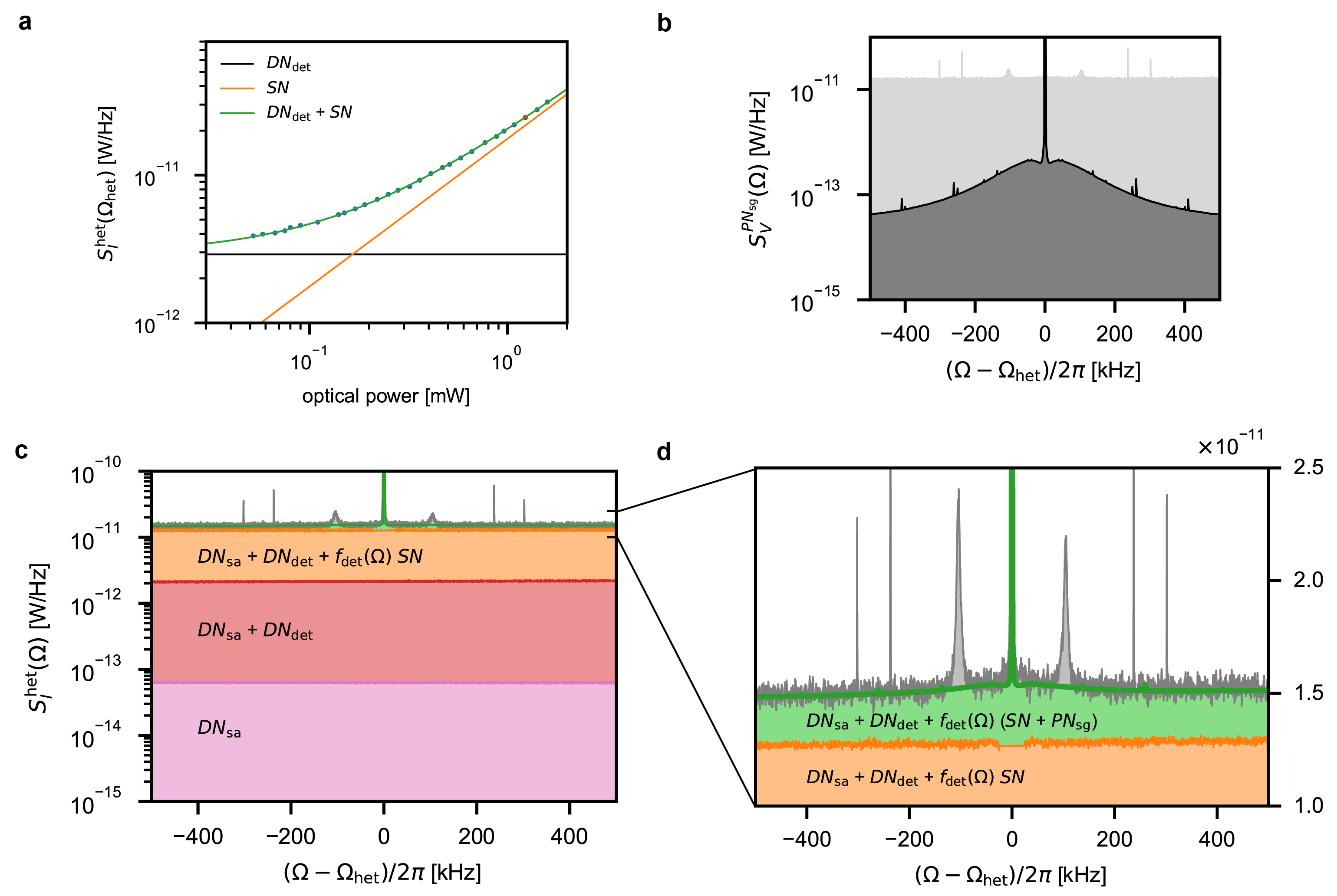}
\caption{\textbf{Noise components in the heterodyne spectra}. \textbf{a} Linear dependence of the shot-noise power as a function of optical power in the heterodyne local oscillator. The red point shows our operating condition, almost a factor 10 above dark noise. \textbf{b} Phase noise of the heterodyne signal generators, directly measured after a mixer, and renormalized to the optical carrier amplitude. The lighter background shows comparison with the raw optical signal. Even though at the relevant frequencies this is almost a factor 100 smaller than the measured signal, its contribution is fundamental (green area in \textbf{c} and \textbf{d}) given low scattering rates in the ground state. \textbf{c} and \textbf{d} Detail of all of the noises contributing to the heterodyne spectrum.}
\label{fig:hetnoise}
\end{figure*}

\begin{figure*}[!htb]
\includegraphics[scale=1]{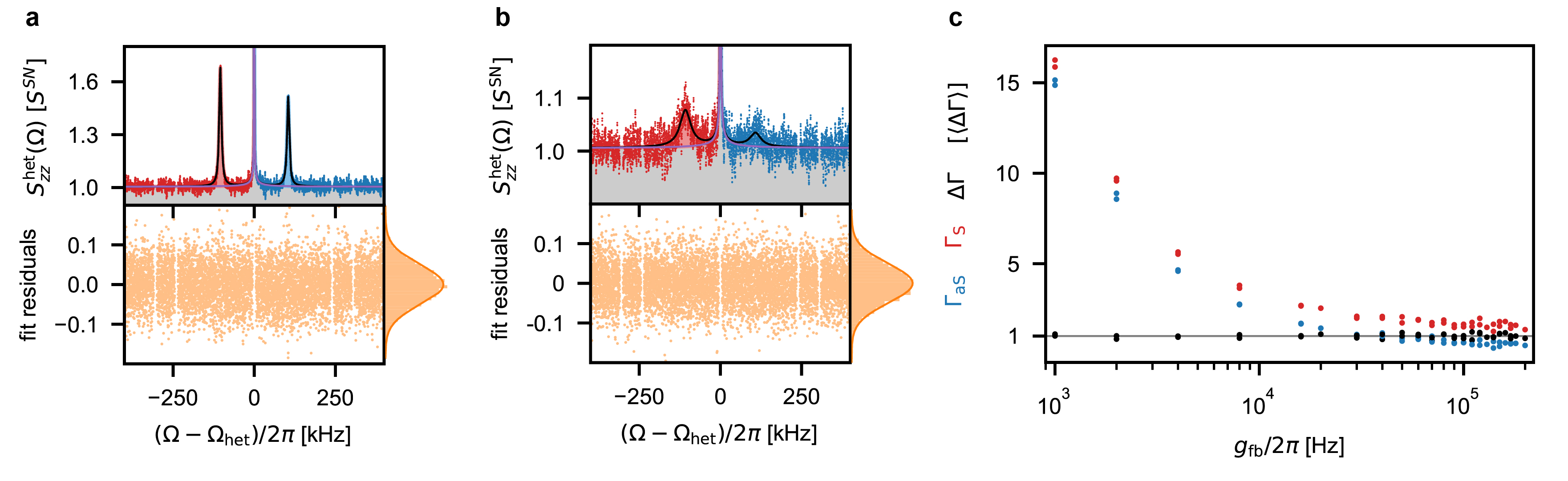}
\caption{\textbf{Side-band asymmetry fit}. \textbf{a} and \textbf{b}, Heterodyne spectra after noise subtraction, application of the inverse detector transfer function $f^{-1}_\mathrm{det}(\omega)$ (whitening). The fit (black line) is a 4 parameter fit of a double lorentian plus a $1/f$ symmetric noise (shown also separately as purple line and gray area). The Gaussian distribution of the residuals shows good agreement of the measured spectra to the noise model. \textbf{c}, Power of the red and blue sideband normalized by their average difference, as a function of the LQG feedback gain for both positive and negative heterodyne frequencies (2 points per colour per gain). The constant difference in the power of the 2 side-bands represents a sanity check of the noise analysis.}
\label{fig:hetfit}
\end{figure*}

By identification of all noise sources in our heterodyne measurement, we are able to isolate the signal component originating from the optomechanical interaction. From that we can evaluate the asymmetry of the Stokes and anti-Stokes peaks.
The heterodyne local oscillator is generated by a sequence of two acousto-optical modulators (AOMs) driven by two locked signal generators at $205\, \mathrm{MHz}$ and $195.8$ (or $214.2$) MHz, aligned to order -1 and +1 respectively, in order to produce a local oscillator shifted by $-9.2$ (or $+9.2$) MHz.
The noise contributions in the heterodyne spectra are determined by: the spectrum analyzer dark noise ($DN_\mathrm{sa}$), the detector dark noise ($DN_\mathrm{det}$), the optical shot noise ($SN$) (Figure \ref{fig:hetnoise} a), the heterodyne signal generator phase noise ($PN_\mathrm{sg}$) and finally the particle's motional signal ($SIG$). In addition one has to also consider the detector transfer function $f_\mathrm{det}(\Omega)$, arising from the 75 MHz cut off frequency. The total noise is 
\begin{equation}
S_\mathrm{raw} = DN_\mathrm{sa} + DN_\mathrm{det} + f_\mathrm{det}(\Omega) (SN + PN_\mathrm{sg} + SIG)  
\end{equation}
Switching on the noise contributions one by one, we are able to directly measure their progressive sum, and evaluate the contribution of each component (Figure \ref{fig:hetnoise}c). As the optical shot noise is white by definition, we can evaluate the detector transfer function (linear in a band of 1 MHz around the heterodyne frequency) by measuring the detector's response to this white noise. Next we want to characterize the $PN_\mathrm{sg}$, which only appears in the heterodyne measurement together with the motioal sidebands. We thus evaluate this noise source ($PN_\mathrm{sg}$) directly, by mixing the signals driving the AOMs and rescaling the carrier peak to that measured in the optical  heterodyne measurement (see Figure \ref{fig:hetnoise} b).
This contribution is then transformed by $f_\mathrm{det}(\Omega)$ and added to the total noise (green component in Figure \ref{fig:hetnoise} c, d ). The sum of separately evaluated noise contributions fits very well to the raw measured data.
We can now isolate the signal of interest. We note that the signal generator phase noise is not the only source of phase noise into our heterodyne. After subtraction of all independently characterized noise contributions, and normalization to shot noise, we notice a residual noise contribution falling off as $1/f$. This noise is compatible to what we expected from the laser phase noise in our unbalanced ($\sim1\,\mathrm{m}$) interferometer. We fit to the clean spectra the sum of a double lorentian \eqref{eq:heterodyne}, and a symmetric $1/f$ noise component, with a fixed offset of 1. For each fit we evaluate the quality of the model by checking the Gaussianity of the residuals (Fig \ref{fig:hetfit} a, b). In addition we verify quantum consistency by noting that, while the ratio (asymmetry) of blue to red side-band changes as cooling improves, their difference, remains constant (Figure \ref{fig:hetfit}c).   
In order to acquire higher statistical significance, we perform repeated measurements for a subset of points, and extract the asymmetry and occupation from the mean value of red and blue side-band powers (Figure 1b in the main text)~\cite{Suhdir2017}.
In order to exclude any other source of uncorrelated noise that may be altering the observed asymmetry we perform our measurements at  both $\Omega_\mathrm{het} = \pm 9.2\,\mathrm{MHz}$. Except from the swapping of the Stokes and anti-Stokes sidebands in the spectra, we observe no difference in the ratio or in the absolute difference of the scattering rates, confirming correct identification of all of the significant noise sources.

\end{document}